\documentclass[aps,pre,twocolumn,showpacs,floatfix,superscriptaddress]{revtex4}
\usepackage{graphicx}
\usepackage{amsmath}
\usepackage{bm}

\begin{document}

\title{Gauge-invariant electromagnetic response of a chiral \boldmath
$p_x+ip_y$ superconductor}

\author{Roman M. Lutchyn}
\affiliation{Joint Quantum Institute, Department of Physics,
University of Maryland, College Park, Maryland 20742-4111, USA}
\affiliation{Condensed Matter Theory Center, Department of Physics,
University of Maryland, College Park, Maryland 20742-4111, USA}

\author{Pavel Nagornykh}
\affiliation{Joint Quantum Institute, Department of Physics,
University of Maryland, College Park, Maryland 20742-4111, USA}

\author{Victor M.~Yakovenko}
\affiliation{Joint Quantum Institute, Department of Physics,
University of Maryland, College Park, Maryland 20742-4111, USA}
\affiliation{Center for Nanophysics and Advanced Materials,
Department of Physics, University of Maryland, College Park,
Maryland 20742-4111, USA}

\date{version ZK, edited by PN 05/05/08, compiled \today}

\begin{abstract}
  We present a gauge-invariant theory of the electromagnetic response
  of a chiral $p_x+ip_y$ superconductor in the clean limit.  Due to
  the spontaneously broken time-reversal symmetry, the effective
  action of the system contains an anomalous term not present in
  conventional superconductors. As a result, the electromagnetic
  charge and current responses contain anomalous terms, which explicitly
  depend on the chirality of the superconducting order parameter.
  These terms lead to a number of unusual effects, such as coupling of
  the transverse currents to the collective plasma oscillations and a
  possibility of inducing the charge density by the magnetic field
  perpendicular to the conducting planes.  We calculate the
  antisymmetric part of the conductivity tensor (the intrinsic Hall
  conductivity) and show that it depends on the wave vector of the
  electromagnetic field.  We also show that the Mermin-Muzikar
  magnetization current and the Hall conductivity are strongly
  suppressed at high frequencies.  Finally, we discuss the implications of
  the theory to the experiments in $\rm Sr_2RuO_4$.
\end{abstract}

\pacs{
74.70.Pq, 
78.20.Ls, 
74.25.Nf, 
73.43.Cd  
}
\maketitle

\section{Introduction}
\label{Sec:Intro}

Unconventional superconductors with spontaneously broken
time-reversal-symmetry (TRS) recently attracted significant interest
~\cite{Rice, Day}.  The idea of a superconducting pairing violating
additional symmetries of the normal phase (on top of the gauge
symmetry) is intriguing, and there has been a lot of effort to find
materials exhibiting such pairings.  Considerable evidence indicates
that $\rm Sr_2RuO_4$ is an unconventional superconductor with broken
TRS~\cite{Xia06, Kidwingira06, Luke98, Luke00}.  The most convincing
indication is the recent observation of the polar Kerr effect (PKE)
in the superconducting state of $\rm Sr_2RuO_4$ by Xia {\it et.\
al.}~\cite{Xia06}.  In this experiment, a linearly polarized light,
incident along the $z$ direction perpendicular to the conducting
planes of $\rm Sr_2RuO_4$, experiences normal reflection
\cite{quarter-wave}.  It was found in Ref.\ \cite{Xia06} that the
polarization plane of the reflected light is rotated relative to the
polarization plane of the incident light by the Kerr angle
$\theta_K\!\sim \! 65$ nrad.  The effect appears below the
superconducting transition temperature $T_c=1.5$ K.  The Kerr
rotation, which may be clockwise or counterclockwise, develops in
the absence of an external magnetic field and is a clear signature
of the spontaneous TRS breaking in the superconducting state.  The
experiment \cite{Xia06} used the Sagnac interferometer, where two
counterpropagating laser beams retrace their paths, so that all
effects, other than the TRS breaking in the sample, cancel out
exactly.  Although previous muon-spin-relaxation measurements
\cite{Luke98, Luke00} gave an indirect evidence that the TRS is
broken in $\rm Sr_2RuO_4$, the PKE experiment \cite{Xia06} provides
much stronger evidence for this remarkable effect.  Additional
indication for the TRS breaking in $\rm Sr_2RuO_4$ comes from the
Josephson junction experiment in the presence of a magnetic field
\cite{Kidwingira06}, which was interpreted as the evidence for
existence of domains with opposite chiralities.  On the other hand,
the scanning SQUID and Hall probe experiments~\cite{Moller05,
Moller07}, designed to search for domains with opposite chiralities
at the surface of $\rm Sr_2RuO_4$, did not find any evidence for the
TRS breaking.  These results show that macroscopic manifestations of
the microscopic TRS breaking are not fully understood and require
further theoretical investigation.  In this paper, we study the
electromagnetic properties of a chiral superconductor with the
$p_x+ip_y$ pairing in the clean limit.

$\rm Sr_2RuO_4$ is a layered perovskite material consisting of
weakly coupled two-dimensional (2D) metallic sheets parallel to the
$(x,y)$ plane \cite{Mackenzie03,Bergemann03}.  It was proposed
theoretically that the superconducting pairing in this material is
spin-triplet \cite{Baskaran96} and has the chiral $p_x+ip_y$ or
$p_x-ip_y$ symmetry \cite{Rice95}.  In this state, Cooper pairs have
the orbital angular momentum $L_z=+\hbar$ or $L_z=-\hbar$ normal to
the layers. Such an order parameter breaks the TRS and is analogous
to the 2D superfluid $^3$He-A \cite{Volovik88}.  It should be
emphasized that the questions of the spin symmetry (singlet vs
triplet) and the orbital symmetry (chiral vs nonchiral) are separate
issues.  It is possible to construct chiral order parameters for
both triplet and singlet pairing \cite{Mazin05}. There is
substantial experimental evidence in favor of the spin-triplet and
odd orbital symmetry of pairing in $\rm Sr_2RuO_4$
\cite{Mackenzie03}, which includes measurements of the spin
susceptibility \cite{Ishida98,Ishida01,Murakawa04} and the Josephson
effect \cite{Nelson04}.  However, there are also alternative
interpretations \cite{Mazin05} in terms of singlet pairing.  We
study the electromagnetic response of quasi-2D (Q2D) chiral
superconductors, and our results are applicable (with minor
modifications) for either spin symmetry.  We do not pay special
attention to the existence of nodal lines in the order parameter of
$\rm Sr_2RuO_4$ (see, e.g.,\ Ref.\ \cite{Sengupta02} for an
interpretation of tunneling measurements). The nodal lines do not
affect the chiral response qualitatively, so we concentrate on the
simplest case of the $p_x+ip_y$ pairing.

Although the $p_x+ip_y$ superconducting pairing breaks the TRS and,
in principle, permits a non-zero Kerr angle $\theta_K$, an explicit
theoretical calculation of $\theta_K$ is challenging.  A textbook
formula \cite{White-Geballe} expresses $\theta_K$ in terms of the ac
Hall conductivity $\sigma_{xy}(\omega)$ at the optical frequency
$\omega$.  A calculation of the intrinsic $\sigma_{xy}(\omega)$ for
a chiral superconductor in the absence of an external magnetic field
turns out to be quite nontrivial.  It is customary for theoretical
calculations to use the gauge where the scalar potential $A_0$ is
set to zero and only the vector potential $\bm A$ is considered.  In
this gauge, the calculations \cite{Joynt91,Ting07} show that there
are no chiral terms in the single-particle response of a $p_x+ip_y$
superconductor.  Using this gauge and taking into account
particle-hole asymmetry, Ref.\ \cite{Yip92} found a small chiral
response from the collective flapping modes.

However, when calculations are performed in a general gauge, they do
produce a non-trivial Chern-Simons-type~(CS) term in the effective
action of a Q2D $p_x+ip_y$ superconductor, which breaks the TRS:
\begin{equation}
  S_{\rm CS} = \Theta \! \int dt\,d^3r\,
  (A_0+\partial_t\Phi/2e)\,(\partial_yA_x-\partial_xA_y)/c.
\label{CS}
\end{equation}
Here, $\Phi$ is the phase of the superconducting order parameter;
$e$ and $c$ are the electron charge and the speed of light,
respectively. Equation\ (\ref{CS}) was first derived in
Ref.~\cite{Volovik88} at $T=0$, and the coefficient $\Theta$ was
found to be $\Theta=\pm e^2/2hd$, where $h$ and $d$ are the Planck
constant and the distance between the layers. The signs $\pm$
correspond to the $p_x\pm ip_y$ pairing.  The term (\ref{CS}) was
then studied in Refs.\
\cite{Goryo98,Goryo99,Goryo00,Golub03,Stone04,Furusaki01}, either at
$T=0$ or at $T$ close to $T_c$ using the Ginzburg-Landau expansion.
In a recent paper \cite{Yakovenko07}, $S_{\rm CS}$ was obtained for
a finite frequency $\omega$ and arbitrary temperature $T$, in which
case Eq.\ (\ref{CS}) should be written in the Fourier representation
with the coefficient $\Theta(\omega)$ under the integral.  Ref.\
\cite{Yakovenko07} found that $\Theta(\omega)$ decreases as
$(\Delta_0/\hbar\omega)^2$ at high frequencies
$\hbar\omega\gg\Delta_0$, where $\Delta_0$ is the superconducting
gap [see Eqs.\ (\ref{Theta})--(\ref{asymptotes})].

Equation\ (\ref{CS}) is the only term in the effective action of a
$p_x+ip_y$ superconductor that breaks the TRS.  Under the
time-reversal operation, the variables in Eq.\ (\ref{CS}) transform
as $\bm A\to-\bm A $, $A_0\to A_0$, $\partial_t\to-\partial_t$, and
$\Phi\to-\Phi$, so $S_{\rm CS}$ changes sign.  The coefficient
$\Theta$ in Eq.\ (\ref{CS}) explicitly depends on the chirality of
the order parameter $p_x\pm ip_y$ and changes sign when the
time-reversal operation is applied to the superconducting pairing
itself.  The action $S_{\rm CS}$ is similar to the standard
Chern-Simons term, which has the structure
$\eta_{\mu\nu\lambda}A_\mu\partial_\nu A_\lambda$, with
$\eta_{\mu\nu\lambda}$ being the (2+1)D antisymmetric tensor and the
indices $\mu$, $\nu$, and $\lambda$ taking the values $t$, $x$, and
$y$.  Compared with the full, gauge-invariant Chern-Simons action,
Eq.~(\ref{CS}) misses the term $A_x\partial_tA_y-A_y\partial_tA_x$.
In Eq.\ (\ref{CS}), the gauge invariance is ensured by the presence
of the superconducting phase $\Phi$, which changes upon a gauge
transformation to compensate for the change of $A_0$
\cite{Volovik88}.

Taking a variation of the effective action with respect to $\bm A$,
one obtains the electric current $\bm j=-c\delta S/\delta\bm A$.  For a
full Chern-Simons term, this would give the usual Hall effect
$j_x=\sigma_{xy}E_y$.  However, a variation of Eq.\ (\ref{CS}) gives
the following (anomalous) current
\begin{equation}
  \bm j^{(a)} =- \Theta \,\hat{\bm z} \times
  \bm\nabla (A_0 + \partial_t\Phi/2e),
\label{j_xy}
\end{equation}
where $\hat {\bm z}$ is a unit vector perpendicular to the
conducting planes.  It was shown in Ref.\ \cite{Stone04} that the
current (\ref{j_xy}) can be expressed as a Mermin-Muzikar current
\cite{Mermin80}
\begin{equation}
  \bm j^{(a)} =  \frac{\hbar}{4m_e} \,
  \hat{\bm z} \times \bm\nabla\rho,
\label{j_M}
\end{equation}
where $\rho$ is the electron charge density, and $m_e$ is the
electron mass.  Equation (\ref{j_M}) can be understood as the
magnetization current $\bm j=\bm\nabla\times\bm M$ originating from
the magnetization $\bm M=-(\hbar\rho/4m_e)\hat{\bm z}$ associated
with the angular momentum $\bm L=\hbar\hat{\bm z}$ of each Copper
pair \cite{Stone04,Mermin80}.  While Eq.\ (\ref{j_M}) is valid at
low frequencies, our calculations show that at high frequencies it
is suppressed by a factor $(\Delta_0/\hbar\omega)^2$, see also Ref.\
\cite{Mineev07}. Moreover, we show that, in addition to the
magnetization current (\ref{j_M}), there is also an anomalous
electric polarization current, that gives a comparable contribution
even at low frequencies.

Equation (\ref{j_xy}) is similar to the standard expression for the
Hall conductivity, but its right-hand side does not contain the
complete electric field $\bm E=-\bm\nabla A_0-\partial_t\bm A/c$ and
includes the superconducting phase $\Phi$.  Equation (\ref{j_xy}) can
be expressed in terms of $\bm E$ as
\begin{equation}
  \bm j^{(a)} = \Theta \hat {\bm z} \times \left(\bm E
  -\frac{\partial\,[\bm\nabla\Phi-(2e/c)\bm A]}{2e\,\partial t}
  \right).
\label{j_xy'}
\end{equation}
The second term in the brackets of Eq.\ (\ref{j_xy'}) is proportional
to the acceleration of the London supercurrent $\bm
j_s=(\rho_s/2m_e)[\bm\nabla\Phi-(2e/c)\bm A]$, where $\rho_s$ is the
superfluid charge density.  It was argued in Ref.\ \cite{Yakovenko07}
that this term in Eq.\ (\ref{j_xy'}) becomes ineffective at high
frequencies and may be neglected.  Then, Eq.\ (\ref{j_xy'}) reduces to
the conventional Hall relation, and the coefficient $\Theta$ can be
identified with the Hall conductivity $\sigma_{xy}=\Theta$ and used
for a calculation of the Kerr angle $\theta_K$
\cite{Yakovenko07,Mineev07}.

However, within the two-fluid model of superconductivity, one can
argue that Newton's equation of motion for the supercurrent is
$(m/e)\,\partial_t\bm j_s=\rho_s\bm E$, i.e.,\ the supercurrent is
accelerated by the electric force.  Then, the right-hand side of
Eq.\ (\ref{j_xy'}) vanishes, and the chiral Hall current $\bm
j^{(a)}$ is zero~\cite{Kallin}.  The reason for this cancellation is
that the superconducting phase $\Phi$ has its own dynamics and
compensates the electromagnetic field in Eq.~(\ref{j_xy})
\cite{FISDW,Yakovenko96,Goan98}. In general, Eq.\ (\ref{j_xy})
should be supplemented with an equation of motion for $\Phi$, and
then $\Phi$ should be eliminated, so that the current response is
expressed in terms of the electromagnetic field only.  In other
words, one should derive the effective action $S(A_\mu,\Phi)$ for a
$p_x+ip_y$ superconductor as a function of the electromagnetic field
$A_\mu$ and the superconducting phase $\Phi$, and then integrate out
$\Phi$ and obtain a new action $S(A_\mu)$ in terms of the
electromagnetic field only.  This procedure is well established for
nonchiral superconductors \cite{AES,Otterlo,Sharapov,Paramekanti},
and it was implemented for chiral superfluids in Refs.\
\cite{Goryo98,Goryo99,Golub03}. However, the calculations in Refs.\
\cite{Goryo98,Goryo99,Golub03} were performed only in the limit of
low frequencies.  As a result, some terms in the effective action
were neglected, and the frequency dependence of the coefficients in
the action was not considered. In our paper, we perform a detailed
derivation of the effective action by taking into account dynamics
of $\Phi$ and the internal Coulomb potential.  Our results are
applicable for all frequencies and exhibit non-trivial frequency
dependence.

Our calculations show that the effective Hall conductivity depends
not only on frequency, but also on the wave vector and is
proportional to the square of the wave vector $q_\|^2=q_x^2+q_y^2$
parallel to the layers: $\sigma_{xy}\propto q_\|^2$.  This
conclusion agrees with Refs.\ \cite{Goryo98,Goryo99,Golub03}.  By
taking the limit of $\bm q\to0$ at $\omega\neq0$ \cite{Mahan}, we
find that the Hall conductivity for a spatially homogeneous system
vanishes, i.e.,\ the cancellation in Eq.~(\ref{j_xy'}) indeed takes
place.  This result is consistent with the general conclusion of
Ref.~\cite{ReadGreen}, which argued that an electric field cannot
produce a sideways motion of the electron gas without an external
magnetic field, no matter what the internal interaction between
electrons producing the $p$-wave pairing is. This argument is based
on the Galilean invariance and is applicable to an infinite,
spatially homogeneous, clean system without boundaries and
impurities.

\begin{figure} \centering
\includegraphics[height=2.75in]{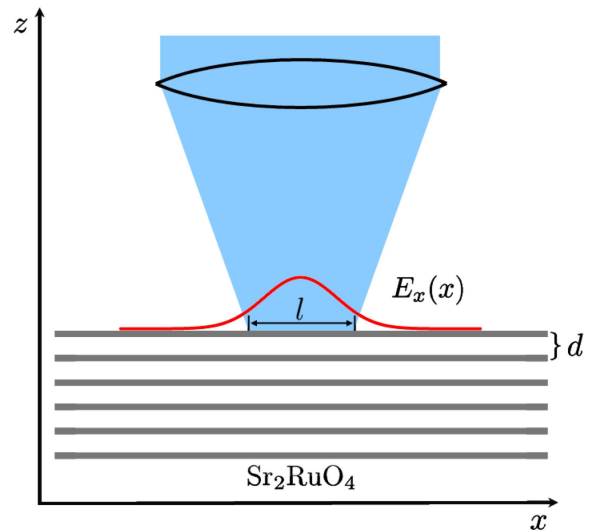}
  \caption{(color online). Schematic picture of the experimental setup
  in Ref.\ \cite{Xia06}.  The incoming laser beam is incident normally
  to the conducting planes of the layered superconductor $\rm
  Sr_2RuO_4$.  The beam is focused by the lens to a spot of the
  diameter $l$ at the surface of the sample.  The solid (red) line
  shows the Gaussian profile of the polarized electric field $E_x$ in
  the beam (Ref.\ \cite{quarter-wave}).  The interlayer distance in $\rm
  Sr_2RuO_4$ is $d$.}
\label{fig:setup}
\end{figure}

This conclusion does not contradict the results of the PKE
experiment, because the setup used in Ref.~\cite{Xia06} generates
spatial inhomogeneities within the ($x,y$) plane of $\rm Sr_2RuO_4$.
Indeed, as sketched in Fig.~\ref{fig:setup}, Xia
\emph{et.~al.}~\cite{Xia06} used a tightly focused Gaussian laser
beam of the transverse size $l\sim25$ $\mu$m, not an infinite
uniform electromagnetic plane wave.  (The beam size $l$ is smaller
than the sample size and the size of a domain with a given
chirality.)  In the Fourier representation, this means that the
electromagnetic wave has non-zero in-plane Fourier components $q_\|$
of the order of $1/l$. Because $\theta_K$ is proportional to
$\sigma_{xy}\propto q_\|^2$, we conclude that the Kerr angle should
be inversely proportional to $l^2$,
\begin{equation}
  \theta_K\propto1/l^2.
\label{size}
\end{equation}
The theoretical prediction (\ref{size}) can be checked experimentally.
It is commonly assumed in literature that $\theta_K$ depends only on
$\omega$ and on the properties of a material \cite{White-Geballe}.
However, Eq.~(\ref{size}) shows the Kerr angle for a chiral
superconductor also depends on the geometrical size of the laser spot.

Equation (\ref{j_M}) shows that the chiral current of a $p_x+ip_y$
superconductor directly couples to charge response and plasma
collective modes.  The Chern-Simons-type term~(\ref{CS}) couples the
longitudinal and transverse electromagnetic fields and leads to a
number of interesting effects, that are not present in nonchiral
superconductors.  By taking a variation of Eq.\ (\ref{CS}) with
respect to $A_0$, we find an anomalous charge response to the
transverse electromagnetic field: $\delta \rho^{(a)}\propto B_z$,
i.e.,\ the charge is induced by the magnetic field $B_z$ along the
$z$ direction \cite{Goryo00,Stone04}.  In Sec.~\ref{Sec:AFM}, we
propose an experimental setup to directly verify this relationship.
We also calculate how the relationship between $\delta\rho^{(a)}$
and $B_z$ is modified at high frequencies.  By the continuity
relation, the induced charge $\delta\rho^{(a)}$ produces an electric
polarization current, which gives an additional contribution to the
Hall effect of the same order as the magnetization current.

The paper is organized as follows.  In Sec.~\ref{Sec:derivation-2D},
we rigorously derive a general expression for the effective action
of a Q2D $p_x+ip_y$ superconductor and obtain the gauge-invariant
electromagnetic response.  In
Secs.~\ref{Sec:Collective}--\ref{Sec:Anomalous}, we discuss the
collective modes and the conventional and anomalous (chiral)
electromagnetic responses.  In Sec.~\ref{Sec:Total}, we obtain the
symmetric and antisymmetric parts of the conductivity tensor of a
chiral superconductor.  In Sec.~\ref{Sec:experiment}, we discuss the
relationship of our results with the experimental studies of the
chiral response of $\rm Sr_2RuO_4$.  Finally, we summarize the
results in Sec.~\ref{Sec:conclusions}.  Some technical details are
relegated to the Appendixes.  In particular, a simplified
alternative derivation of the effective action is given in Appendix
\ref{Sec:alt-action}.

\section{Effective action for a chiral quasi-two-dimensional \boldmath $p_x+ip_y$
superconductor}
\label{Sec:derivation-2D}

\subsection{Triplet \boldmath$p_x+ip_y$ pairing}
\label{Sec:Pairing}

First, we briefly summarize basic information about the triplet
$p$-wave pairing \cite{Leggett75}.  The Cooper pairing between
electrons is described by the pairing potential
$\Delta_{\alpha\beta}(\bm r,\bm r')\propto\langle \psi_\alpha(\bm
r)\psi_\beta(\bm r')\rangle$. Here, $\psi_\alpha(\bm r)$ is the
electron annihilation operator at the point $\bm r$ with the spin
projection $\alpha=\uparrow,\downarrow$.  For a uniform,
translationally-invariant system, the pairing potential depends only
on the relative distance $\bm r-\bm r'$, so one can perform the
Fourier transform and use the momentum representation
$\Delta_{\alpha\beta}(\bm p)\propto\langle \psi_\alpha(\bm
p)\psi_\beta(-\bm p)\rangle$.  The pairing tensor
$\Delta_{\alpha\beta}(\bm p)=\Delta(\bm p)\,\bm d(\bm p)\cdot{\bm
\sigma}_\alpha^\gamma\,\eta_{\gamma\beta}$ can be written in terms
of the antisymmetric metric tensor $\eta_{\gamma\beta}$ and the
Pauli matrices ${\bm\sigma}_\alpha^\gamma$, where the unit vector
$\bm d(\bm p)$ characterizes the spin polarization of the triplet
state.  The prefactor $\Delta(\bm p)$ is a momentum-dependent
pairing amplitude.

For $\rm Sr_2RuO_4$, we consider the case where the vector $\bm
d(\bm p)=\hat {\bm z}$ has the uniform, momentum-independent
orientation, which represents pairing between electrons with the
opposite spins $\langle\psi_\uparrow(\bm p)\psi_\downarrow(-\bm
p)\rangle$.  (It can be transformed into pairing with parallel spins
by changing the spin quantization axis.)  For the orbital symmetry,
we consider the chiral pairing potential $\Delta(\bm
p)=\Delta_0(p_x\pm ip_y)/p_F$, where $p_F$ is the Fermi momentum,
and $\Delta_0$ is the superconducting gap.  This order parameter
corresponds to a vortex in the momentum space, because the phase of
$\Delta(\bm p)$ changes by $\pm2\pi$ when $\bm p$ goes around the
Fermi surface.  It is instructive to write the pairing potential in
the form
\begin{equation}
  \Delta(\bm p)=\Delta_xp_x+i\Delta_yp_y,
\label{Delta_xy}
\end{equation}
and set $\Delta_x=\pm\Delta_y=\Delta_0/p_F$ only at the end of the
calculations.  The sign of the product
\begin{equation}
  s_{xy}\equiv{\rm sign}(\Delta_x\Delta_y)
\label{s_xy}
\end{equation}
reflects the sign of the order-parameter chirality.

\subsection{Theoretical model}
\label{Sec:Model}

Our goal is to derive an effective action for a chiral superconductor
in a weak electromagnetic field.  This approach is equivalent to a
linear response calculation \cite{Schrieffer, UFN, Kulik, Levin, Das
Sarma}.  We use the Greek indices, e.g.,\ $\mu$ and $\nu$, to denote
the space-time components of tensors in the Minkowski notation and the
Roman indices, e.g.,\ $k$ and $l$, for the space components.  To
simplify intermediate steps of calculations, we set the Planck
constant $\hbar$, the Boltzmann constant $k_B$, and the speed of light
$c$ to unity: $\hbar=k_B=c=1$.  The constants $\hbar$ and $k_B$ can be
easily restored in the final equations by dimensionality.  The speed
of light can be restored by noting that it appears only in the
combination $\bm A/c$ with the vector potential $\bm A$.  To simplify
presentation, we first study a purely 2D case, which corresponds to
just one metallic layer in the $(x,y)$ plane, and then generalize the
calculation to the case of many coupled parallel metallic layers, as
appropriate for $\rm Sr_2RuO_4$.

The Hamiltonian of interacting electrons, subject to an external
electromagnetic field $A^{\mu}$, is given by
\begin{eqnarray}
  \hat H &=& \int d^2 r\, \psi^\dag_\sigma(\bm r)
  \left(\frac{[\hat{\bm p}-e\bm A(\bm r)]^2}{2m_e}-\mu\right)
  \psi_\sigma(\bm r)
\nonumber \\
  &+& \int d^2 r\, eA_0(\bm r) \, \delta n(\bm r)
\label{Hamiltonian} \\
  &-& \int d^2 r \, d^2 r' g(\bm r-\bm r')\,
  \psi^\dag_\uparrow(\bm r) \psi^\dag_\downarrow(\bm r')
  \psi_\downarrow(\bm r') \psi_\uparrow(\bm r)
\nonumber \\
  &+& \frac{e^2}{2} \int d^2 r\, d^2 r'
  \delta n(\bm r) \, V(\bm r-\bm r') \, \delta n(\bm r').
\nonumber
\end{eqnarray}
Here, $m_e$ and $\mu$ are the electron mass and the chemical
potential, $g(\bm r-\bm r')$ is an anisotropic interaction potential
leading to a $p$-wave pairing, and $V(\bm r-\bm r')=1/|\bm r-\bm
r'|$ is the Coulomb interaction potential. The density fluctuation
operator $\delta n(\bm r)$ reads
\begin{equation}
  \delta n(\bm r)=\psi^\dag_\sigma(\bm r)\psi_\sigma(\bm r)-n_0,
\end{equation}
with $n_0$ being the 2D background charge density. The summation over
repeated indices is assumed everywhere.

The starting point for our calculation is the partition function $Z$,
which can be expressed as a path integral over the anticommuting
fermionic fields $\psi$ and $\psi^\dag$.  We use the
Hubbard-Stratonovich transformation to decouple the fermion
interaction terms in Eq.\ (\ref{Hamiltonian}) by introducing
additional integrals over auxiliary bosonic fields $\Delta$ and
$\varphi$ \cite{Negele, Svidzinskii, AES, Otterlo, Sharapov}.  The
complex field $\Delta(\bm r,\bm r',\tau)$ is the superconducting
pairing potential, and $\varphi(\bm r,\tau)$ is the internal electric
potential produced by electrons.  As a result, the partition function,
written in the imaginary time $\tau$, reads \cite{Euclidean}
\begin{equation}
  Z = \int \mathcal D\Delta^* \mathcal D\Delta \mathcal D\varphi \:
  e^{-S_{\rm bos}}
  \int \mathcal D\psi^\dag \mathcal D\psi \:
  e^{-S_{\rm el}},
\label{partition_with_A}
\end{equation}
where the bosonic action is
\begin{eqnarray}
  && S_{\rm bos} =
   ie\int d\tau  d^2 r \, [\varphi(\bm r,\tau) + A_0(\bm r,\tau)]\,n_0
\label{L_bos} \\
  && + \int \!d\tau  d^2 r\, d^2 r' \left[
  \frac{|\Delta(\bm r,\bm r',\tau)|^2}{g(\bm r\!-\!\bm r')}
  +\frac{\varphi(\bm r,\tau)\varphi(\bm r',\tau)}{2V(\bm r\!-\!\bm r')}
  \right],
\nonumber
\end{eqnarray}
and the electronic action is
\begin{eqnarray}
  && S_{\rm el} \!=\!
  \int \! d\tau  d^2 r  \psi^\dag_{\sigma}(\bm r,\tau)\!
  \left(\frac{[\hat{\bm p}-e\bm A(\bm r,\tau)]^2}{2m_e}-\mu\right)\!
  \psi_\sigma(\bm r,\tau)
\nonumber \\
  &&\! +\! \int \! d\tau d^2 r \,
  \psi^\dag_{\sigma}(\bm r,\tau)\!
  \{\partial_\tau-ie[\varphi(\bm r,\tau)\!+\!A_0(\bm r,\tau)]\}\!
  \psi_\sigma(\bm r,\tau)
\nonumber \\
  && \!-\! \int \! d\tau d^2 r  d^2 r' \left[\Delta(\bm r,\bm r',\tau)
  \psi^\dag_\uparrow(\bm r,\tau)\psi^\dag_\downarrow(\bm r',\tau)
 \! +\! \mathrm{H.c.} \right]\!\!.
\label{L_el}
\end{eqnarray}
The superconducting pairing potential in Eq.\ (\ref{L_el}) can be
written as a function of the relative coordinate $\bm r-\bm r'$ and
the center-of-mass coordinate $\bm R=(\bm r+\bm r')/2$. For a uniform
system, in the absence of electromagnetic field, the equilibrium
saddle-point configuration of $\Delta_0$ in the total action
(\ref{partition_with_A}) does not depend on $\bm R$ and $\tau$, and is
given by Eq.\ (\ref{Delta_xy}) in the Fourier representation with
respect to $\bm r -\bm r'$.  In the presence of an applied
electromagnetic field, the saddle-point value of the complex
superconducting pairing potential is shifted and can be written in
the adiabatic Born-Oppenheimer approximation \cite{Sharapov, Thouless}
as
\begin{equation}
  \Delta(\bm p,\bm R,\tau)=\frac{p_x+ip_y}{p_F}
  [\Delta_0+\delta\Delta(\bm R,\tau)]\,
  e^{i\Phi(\bm R,\tau)}.
\end{equation}
We assume here that the applied electromagnetic field is weak, and
varies slowly in space and time, i.e.,\ $\xi/\lambda \ll 1$ and
$\omega\ll \omega_{\rm cutoff}< \varepsilon_F$. Here, $\lambda$ and
$\omega$ are the wavelength and frequency of the electromagnetic
field, $\xi=v_F/\Delta_0$ is the superconducting coherence length,
$\omega_{\rm cutoff}$ is the cutoff frequency of the interaction
responsible for superconducting pairing, and $\varepsilon_F$ is the
Fermi energy.  The field shifts the amplitude of the pairing
potential by $\delta\Delta$ and gives a space-time dependence to the
phase $\Phi$ of the order parameter.  As in the $s$-wave
superconductors, the amplitude variations $\delta\Delta$ are massive
\cite{Kulik}, and their contribution to the linear electromagnetic
response is small in the parameter $(\Delta_0/\varepsilon_F)^2\ll 1$
\cite{UFN}.  Therefore, we neglect the amplitude fluctuations in the
rest of the paper and only consider the dynamics of the phase
$\Phi$.  The $p$-wave order parameter also has other modes, such as
the clapping modes, due to its internal orbital structure
\cite{Yip92}. However, these modes are also massive, and we do not
consider them.

The phase $\Phi$ of the order parameter is essential for ensuring the
gauge invariance of the theory.  By performing a unitary
transformation of the fermion operators $\psi_\sigma(\bm r,\tau)
\rightarrow e^{i\Phi(\bm r,\tau)/2}\psi_\sigma(\bm r,\tau)$, one can
compensate the phase of the order parameter \cite{Sharapov}.  As a
result, the electromagnetic field in Eq.\ (\ref{L_el}) is replaced by
the gauge-invariant combinations of electromagnetic potentials and the
superconducting phase:
\begin{equation}
  \tilde A_0 = A_0 - \partial_\tau\Phi/2e, \quad
  \tilde{\bm A} = \bm A - \bm\nabla\Phi/2e.
\label{tilde-A}
\end{equation}
In this way, one ensures that the gauge invariance is fulfilled at
every step of the calculation.

After the phase transformation, the superconducting pairing
potential in Eq.\ (\ref{L_el}) has the equilibrium value given by
Eq.\ (\ref{Delta_xy}).  The electron Lagrangian (\ref{L_el}) can be
written in the momentum-frequency representation as a $2\times2$
Nambu matrix acting on the spinor $\bm\psi(p)=[\psi_\uparrow(\bm
p,\omega_m),\psi_\downarrow^\dag(-\bm p,\omega_m)]$, where
$\omega_m$ is the fermionic Matsubara frequency, and
$p=(i\omega_m,p_x,p_y)$,
\begin{eqnarray}
  && S_{\rm el} = \sum_p \bm\psi^\dag(p)\, G^{-1}(p)\, \bm\psi(p)
\label{lagrangian2} \\
  && + \sum_{p,q} \bm\psi^\dag(p+q) \Gamma_1(p,q)\bm\psi(p)
  + \sum_{p,q} \bm\psi^\dag(p) \Gamma_2(p, q)\bm\psi(p).
\nonumber
\end{eqnarray}
Here, $\sum_p$ represents the integration over momenta as well as
the summation over the Matsubara frequencies \cite{Gamma_2}. The
terms $G^{-1}$, $\Gamma_1$, and $\Gamma_2$ in the electron
action~(\ref{lagrangian2}) contain the zeroth, first, and second
powers of the electromagnetic fields $\tilde A^\mu$ and $\varphi$,
\begin{eqnarray}
  G^{-1}(p) \!\!\!  &=& \!\!\! i\omega_m - \xi_{\bm p}\hat\tau_3
  - p_x\Delta_x\hat\tau_1 + p_y\Delta_y\hat\tau_2,
\label{1/G} \\
  \Gamma_1(p,q) \!\!\! &=& \!\!\! -ie[\tilde A_0(q)+\varphi(q)]\hat\tau_3
  -e \bm v(\bm p+\bm q/2)\cdot \tilde{\bm A}(q),
\label{G_1} \\
  \Gamma_2(p,q) \!\!\!  &=& \!\!\! \frac{e^2}{2}
  \frac{\partial^2 \xi_{\bm p}}{\partial p_k\partial p_l}
  \tilde A_k(q) \tilde A_l(-q)\hat\tau_3.
\label{G_2}
\end{eqnarray}
Here, the wave vector $\bm q$ of the electromagnetic field and its
bosonic Matsubara frequency $\Omega_n$ are combined into
$q=(i\Omega_n,\bm q)$. The electron dispersion and velocity are
$\xi_{\bm p}=\bm p^2/2m_e-\mu$ and $\bm v(\bm p)=\bm p/m_e$.
Equations (\ref{1/G})--(\ref{G_2}) are written in terms of the Pauli
matrices $\hat\tau$ acting on the electron spinor $\bm\psi$.

\subsection{Integrating out fermions}
\label{Sec:Fermions}

Substituting the action (\ref{lagrangian2}) into
Eq.~(\ref{partition_with_A}) and taking the functional integral over
$\psi$, we obtain the electron contribution $S_{\rm eff}^{\rm el}$
to the effective action up to the second order in the
electromagnetic fields $\tilde A^\mu$ and $\varphi$,
\begin{eqnarray}
  && S_{\rm eff}^{\rm el}(\tilde A^\mu,\varphi)
  = -\mathrm{Tr}\ln(G^{-1}+\Gamma_1+\Gamma_2)
  = -\mathrm{Tr}\ln{G^{-1}}
\nonumber \\
  && {} - \mathrm{Tr}\,{G\Gamma_1} - \mathrm{Tr}\,{G\Gamma_2}
  + \frac12\mathrm{Tr}\,{G\Gamma_1G\Gamma_1}.
\label{S_el}
\end{eqnarray}
Here, $\mathrm{Tr}$ denotes both the matrix trace and the sum over
the internal frequencies and momenta.

The term $\mathrm{Tr}\ln{G^{-1}}$ in Eq.\ (\ref{S_el}) and the term
proportional to $|\Delta|^2$ in Eq.\ (\ref{L_bos}) define the saddle
point configuration (\ref{Delta_xy}) of the unperturbed system.  The
functional integral over the amplitude of $\Delta$ is taken by
expansion in the vicinity of this saddle-point
configuration. Combining the remaining terms in Eqs.\ (\ref{L_bos})
and (\ref{S_el}), we obtain the effective phase-only action of the
system
\begin{eqnarray}
  && Z = \int \mathcal D\Phi \, \mathcal D\varphi \,
  \exp[-S_{\rm eff}(\tilde A^\mu,\varphi)],
\label{Z-fermion} \\
  && S_{\rm eff}(\tilde A^\mu,\varphi) = \frac{1}{2}
  \sum_{q}\left[
  \frac{\varphi(q)\varphi(-q)}{V(\bm q)}
  + \frac{e^2n_0}{m} \tilde{\bm A}(q)\cdot \tilde{\bm A}(-q) \right]
\nonumber \\
  && + \frac12 \mathrm{Tr}[G\Gamma_1G\Gamma_1].
\label{S_eff}
\end{eqnarray}
Here, $V(\bm q)$ is the Fourier transform of the Coulomb potential,
which in 2D is equal to $V(\bm q)=2\pi/|\bm q|$.

Now we need to evaluate the trace in Eq.\ (\ref{S_eff}).  This
amounts to calculation of one-loop Feynman diagrams using the
fermionic Green's function $G$ from Eq.\ (\ref{1/G}),
\begin{equation}
  G=-\frac{i\omega_m+\xi_{\bm p}\hat\tau_3+p_x\Delta_x\hat\tau_1
  -p_y\Delta_y\hat\tau_2}
  {\omega_m^2+\xi_{\bm p}^2+p_x^2\Delta_x^2+p_y^2\Delta_y^2}
\label{G},
\end{equation}
and the vertex $\Gamma_1$ (\ref{G_1}), which represents interaction
with the electromagnetic field.  The final expression for the
effective action (\ref{S_eff}) can be written as
\begin{eqnarray}
  && S_{\textrm{eff}}(\tilde A^{\mu},\varphi)
  = \frac{1}{2}\sum_q \frac{\varphi(q)\varphi(-q)}{V(\bm q)}
\label{action1} \\
  && +Q_{00}i[\tilde{A}_0(q)\!+\!\varphi(q)]i[\tilde{A}_0(\!-\!q)\!
  +\!\varphi(\!-\!q)]\!+\!Q_{kl}\tilde{A}_k(q)\tilde{A}_l(\!-\!q)
\nonumber \\
  && + iQ_{0k}[\tilde{A}_0(q)\!+\!\varphi(q)]\tilde{A}_k(\!-\!q)\!
  +\!iQ_{k0}\tilde{A}_k(q)[\tilde{A}_0(\!-\!q)\!+\!\varphi(\!-\!q)].
\nonumber
\end{eqnarray}
Here, $Q_{00}$, $Q_{kl}$, and $Q_{0k}(q)=Q_{k0}(-q)$ are the
corresponding correlation (polarization) functions.

The density-density polarization function
$Q_{00}=e^2{\rm Tr}[\tau_3G\tau_3G]$ is
\begin{equation}
  \!Q_{00}\!=\!\int{\!\frac{d^2 p}{(2\pi)^2}}\!
  \frac{2e^2}{\beta}\sum_{i\omega_m}\!
  \frac{i\omega_m(i\omega_m\!+\!i\Omega_n)\!
  +\!\xi_{-}\xi_{+}\!-\!\Delta_+\Delta_-}
  {[(\omega_m\!+\!\Omega_n)^2\!+\!E_{+}^2][\omega_m^2\!+\!E_{-}^2]}.
\label{Q00}
\end{equation}
Here, $\beta=1/T$, $\xi_\pm=\xi_{\bm p\pm\bm q/2}$, $\Delta_+
 \Delta_-=\Delta_x^+ \Delta_x^- \!+\! \Delta_y^+ \Delta_y^-$,
 $E_\pm=\sqrt{\xi_\pm^2+(\Delta_x^\pm)^2+(\Delta_y^\pm)^2}$ with
 $\Delta_{x(y)}^\pm=(p_{x(y)}\pm q_{x(y)}/2)\Delta_{x(y)}$.

The current-current correlation function $Q_{kl}$ consists of the
diamagnetic $Q^{(1)}_{kl}$ and paramagnetic $Q^{(2)}_{kl}=e^2{\rm
Tr}[v_kGv_lG]$ parts
\begin{eqnarray}
 \!\! Q_{kl} \! &\!=\!& \!  Q^{(1)}_{kl}+Q^{(2)}_{kl}, \quad\quad\quad
 \!\! Q^{(1)}_{kl}=\frac{e^2 n_0}{m_e}\,\delta_{kl},
\label{Qij} \\
  Q^{(2)}_{kl} \!\! &\!\!=\!\!& \!\! \int \!\!
  \frac{d^2p}{(2\pi)^2} \frac{2e^2 v_kv_l}{\beta}\!\sum_{i\omega_m}\!
  \frac{i\omega_m(i\omega_m\!+\!i\Omega_n)\!+\!\xi_{-}\xi_{+}\!
  +\! \Delta_+\Delta_-}
  {[(\omega_m\!+\!\Omega_n)^2\!+\!E_{+}^2][\omega_m^2\!+\!E_{-}^2]},
\nonumber \\
\label{Qij1}
\end{eqnarray}
with $\bm v=\partial \xi_{\bm p}/\partial\bm p=\bm p/m_e$ being the
electron velocity.

The expressions (\ref{Q00}), (\ref{Qij}) and (\ref{Qij1}) for the
density-density and current-current correlation functions are the
same for chiral and nonchiral superconductors and contain nothing
special. The only important difference between chiral and nonchiral
superconductors appears in the structure of the current-density
correlation function $Q_{k0}=e^2{\rm Tr}[v_kG\tau_3G]$.  This
difference plays the crucial role for the results of our paper.  In
a $p_x+ip_y$ superconductor, $Q_{k0}$ consists of the conventional,
symmetric part $Q_{k0}^{(s)}$ and the anomalous, antisymmetric part
$Q_{k0}^{(a)}$:
\begin{eqnarray}\label{Q0j}
  Q_{k0} \!&\!\!=\!\!& Q_{k0}^{(s)}+Q_{k0}^{(a)},
\label{Qj0} \\
  Q_{k0}^{(s)}\!&\!\!=\!\!&\!-\!\int{\frac{d^2 p}{(2\pi)^2}}\!
  \frac{2 e^2 v_k}{\beta} \sum_{i\omega_m}
  \frac{(i\omega_m\!+\!i\Omega_n)\xi_-\!+\!i\omega_m\xi_+}
  {[(\omega_m\!+\!\Omega_n)^2\!+\!E_+^2][\omega_m^2\!+\!E_-^2]},
\nonumber\\ \label{Qj0s}\\
  Q_{k0}^{(a)}\!&\!\!=\!\!&\!-\!\int{\frac{d^2 p}{(2\pi)^2}}\!
  \frac{2 e^2 v_k}{\beta} \sum_{i\omega_m}
  \frac{[iq_xp_y\!-\!iq_yp_x]\Delta_x\Delta_y}
  {[(\omega_m\!+\!\Omega_n)^2\!+\!E_+^2][\omega_m^2\!+\!E_-^2]}.
\nonumber\\ \label{Qj0a}
\end{eqnarray}
The two terms have the following symmetries
\begin{equation}
  Q_{k0}^{(s)}(-q) = Q_{k0}^{(s)}(q), \quad
  Q_{k0}^{(a)}(-q) = -Q_{k0}^{(a)}(q),
\label{s-a}
\end{equation}
where the operation $q\to-q$ means changing the signs of both
frequency and momentum. Equation\ (\ref{s-a}) follows from an
observation that $ Q_{k0}^{(s)}$ is proportional to a product of the
frequency and momentum components of $q$, whereas $ Q_{k0}^{(a)}$ is a
linear function of the momentum components $q_x$ and $q_y$ only. One
can also check using Eqs.\ (\ref{Q00}), (\ref{Qij}) and (\ref{Qij1})
that $Q_{00}(-q)=Q_{00}(q)$ and $Q_{kl}(-q)=Q_{kl}(q)$ are symmetric.

Equation (\ref{Qj0a}) shows that the anomalous charge-current
correlation function $Q_{k0}^{(a)}$ explicitly depends on the
chirality (\ref{s_xy}) of the order parameter, whereas the other
correlation functions do not depend on it.  We demonstrate in the
rest of the paper that all of the non-trivial, chiral response of a
$p_x+ip_y$ superconductor originates from the anomalous term
(\ref{Qj0a}).  The density-current correlator $Q_{k0}$ (\ref{Qj0})
is rarely discussed in textbooks and literature.  Non-chiral
superconductors have only the symmetric term $Q_{k0}^{(s)}$
(\ref{Qj0s}), which couples to the longitudinal degrees of freedom
(see Sec.~\ref{Sec:Q_mu_nu}), such as plasmon collective modes, but
does not affect the transverse response, such as the London-Meissner
current.  Therefore, for the calculation of the transverse response,
it is sufficient to consider a gauge with $\bm A^{\perp}\neq0$ and
$A_0=0$.  However, in chiral superconductors, the antisymmetric term
$Q_{k0}^{(a)}$ couples to the transverse response (see
Sec.~\ref{Sec:Q_k0-a}) and controls the TRS-breaking response of the
system.  In the technical language, when calculating the transverse
response of the chiral superconductors, one has to include vertex
corrections in order to obtain correct results. Without vertex
corrections, calculations do not generate chiral terms in the
electromagnetic response of a $p_x+ip_y$ superconductor
\cite{Joynt91,Ting07}.

\subsection{Integrating out the internal Coulomb potential}
\label{Sec:Coulomb}

\begin{figure}
\centering
\includegraphics[width=0.9\linewidth]{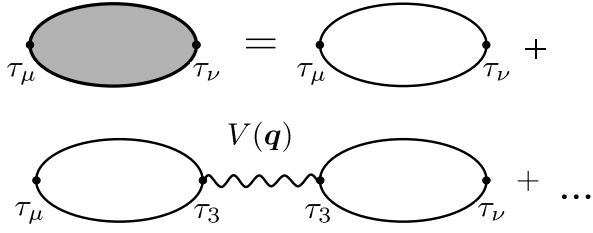}
\caption{Renormalization of the polarization functions (\ref{tilde_Q})
  due to the Coulomb interaction $V(\bm q)$ in the RPA.
  The wavy line denotes $V(\bm q)$, which couples only to the charge
  vertices represented by the Pauli matrices $\tau_3$.}
\label{fig_diag}
\end{figure}

It is well-known that response functions of a charged
superconductor, as opposed to a neutral superfluid such as $^3$He,
are strongly modified by the Coulomb interaction.  In our approach,
integrating out the internal electric potential $\varphi$ is
equivalent to taking into account the Coulomb interaction, because
$\varphi$ mediates the electrostatic interaction between electrons.
The effective action $S_{\rm eff}(\tilde A^\mu)$, obtained after
taking the functional integral over $\varphi$ in
Eq.~(\ref{Z-fermion}), is defined as follows:
\begin{equation}
  Z = \int \mathcal D\Phi \, \mathcal D\varphi \,
  e^{-S_{\rm eff}(\tilde A^\mu,\varphi)}
  = \int \mathcal D\Phi \, e^{-S_{\rm eff}(\tilde A^\mu)}.
\label{Z-Coulomb}
\end{equation}
Substituting Eq.\ (\ref{action1}) into Eq.\ (\ref{Z-Coulomb}) and
taking the Gaussian integral over $\varphi$, we find
\begin{eqnarray}
  S_{\rm eff}(\tilde A^\mu)\!&=&\!\frac{1}{2}
  \sum_q \tilde{Q}_{00}i\tilde{A}_0(q)i\tilde{A}_0(-q)
  +\tilde{Q}_{kl}\tilde{A}_k(q)\tilde{A}_l(-q)
\nonumber \\
  &+&2\tilde{Q}_{0k} i\tilde{A}_0(q)\tilde{A}_k(-q),
\label{action2}
\end{eqnarray}
where the polarization functions are renormalized by the Coulomb
interaction $V(\bm q)$ as follows:
\begin{eqnarray}
  && \tilde{Q}_{00}=\frac{Q_{00}(q)}{1-V(\bm q)\,Q_{00}(q)}, \quad
  \tilde{Q}_{k0}=\frac{Q_{k0}(q)}{1-V(\bm q)\,Q_{00}(q)},
\nonumber \\
  && \tilde{Q}_{kl}=Q_{kl}
  +\frac{Q_{k0}V(\bm q)\,Q_{0l}}{1-V(\bm q)\,Q_{00}}.
\label{tilde_Q}
\end{eqnarray}
The renormalized polarization functions (\ref{tilde_Q}) can be
equivalently obtained through the resummation of the most diverging
diagrams due to the Coulomb interaction in the random phase
approximation~(RPA) as shown in Fig.~\ref{fig_diag}.

\subsection{Integrating out the superconducting phase}
\label{Sec:phase}

We integrate out the superconducting phase $\Phi$ to obtain the
gauge-invariant effective action $S_{\rm eff}(A^\mu)$ that depends
only on the electromagnetic field $A^\mu$ as follows:
\begin{equation}
  Z =  e^{-S_{\rm eff}(A^\mu)} =
  \int \mathcal D\Phi \, e^{-S_{\rm eff}(\tilde A^\mu)},
\label{Z-phase}
\end{equation}
where the field $\tilde A^\mu$ is defined by Eq.~(\ref{tilde-A}) and
contains $\Phi$.  Substituting Eq.~(\ref{action2}) into
Eq.~(\ref{Z-phase}) and taking the Gaussian integral over $\Phi$, we
finally obtain the effective action \cite{Kosztin}
\begin{eqnarray}
  S_{\rm eff}(A^\mu) &=& \frac{1}{2}
  \sum_q K_{\mu\nu}(q) A^\mu(q) A^\nu(-q),
\label{S(A)} \\
  K_{\mu\nu} &=& \tilde Q_{\mu\nu} -
  \frac{\tilde Q_{\mu\rho}q^\rho q^\sigma \tilde Q_{\sigma\nu}}
  {\tilde Q_{\alpha\beta}q^\alpha q^\beta}.
\label{K}
\end{eqnarray}
Notice that the kernel (\ref{K}) satisfies the identity
\begin{equation}
  q^\mu K_{\mu\nu} = K_{\mu\nu} q^\nu = 0,
\label{qK}
\end{equation}
which ensures that the effective action (\ref{S(A)}) is
gauge-invariant.  Indeed, in the momentum representation, the gauge
transformation is $A^\mu \to A^\mu -i q^\mu\phi$, where $\phi$ is an
arbitrary scalar function.  Substituting this expression into Eq.\
(\ref{S(A)}) and taking into account the identities (\ref{qK}), we
see that the additional terms $q^\mu\phi$ drop out and, thus, the
effective action is gauge-invariant.

\subsection{Generalization to the quasi-two-dimensional case}
\label{Sec:Q2D}

The derivation presented in the previous sections can be easily
generalized to a Q2D system consisting of parallel superconducting
layers separated by the distance $d$ in the $\hat{ \bm z}$ direction.
The layers are coupled via the interlayer electron tunneling amplitude
$t_c$.  In the tight-binding approximation, the electron dispersion
and velocity are
\begin{equation}
  \xi_{\bm p} = \frac{p_x^2+p_y^2}{2m_e} - t_c\cos(p_z d) - \mu,
  \quad \bm v(\bm p)=\frac{\partial\xi_{\bm p}}{\partial\bm p},
\label{spectrum3d}
\end{equation}
where $(p_x,p_y)$ and $p_z$ are the in-plane and out-of-plane
momenta, respectively.  We assume that the interlayer tunneling
amplitude is much smaller than the in-plane Fermi energy:
$t_c\ll\varepsilon_F$.  Then, the Fermi surface is a slightly warped
cylinder extended in the $p_z$ direction.  Equation
(\ref{spectrum3d}) should be substituted into Eqs.\
(\ref{Q00})--(\ref{Qj0a}), where the integrals should be taken over
a three-dimensional~(3D) momentum $\bm p$, with $p_z$ limited to the
interval $[-\pi/d,\pi/d]$.

The diamagnetic term $Q^{(1)}_{kj}$ in Eq.\ (\ref{Qij}) should be
replaced by the expression following from Eq.\
(\ref{G_2})~\cite{Sharapov}
\begin{equation}
  Q^{(1)}_{kl} \!=\! e^2 \!\int\! \frac{d^3p}{(2\pi)^3}
  \left[1-\frac{\xi_{\bm p}}{E_{\bm p}}
  \tanh\left(\frac{E_{\bm p}}{2T}\right) \right]
  \frac{\partial^2\xi_{\bm p}}{\partial p_k\,\partial p_l},
\label{Qij13D}
\end{equation}
where $E_{\bm p}=\sqrt{\xi_{\bm p}^2+|\Delta(\bm p)|^2}$. The
integrals (\ref{Qij13D}) can easily be evaluated and give the
following expression for the diamagnetic tensor of a Q2D
superconductor:
\begin{equation}
  Q_{kl}^{(1)}=\frac{e^2 n_0}{m_ed}\,\tensor{n}_{kl}, \quad
  \tensor{n}=
  \left(\begin{array}{ccc}
    1 & 0 & 0 \\
    0 & 1 & 0 \\
    0 & 0 & \displaystyle \frac{t_c^2m_e d^2}{2\varepsilon_F}
  \end{array}\right).
\label{tensor-n}
\end{equation}
Here, $n_0=p_F^2/2\pi$ is the 2D electron density, and $p_F$, $v_F$,
and $\varepsilon_F=p_F^2/2m_e$ are the in-plane Fermi momentum,
velocity, and energy.  The dimensionless tensor $\tensor{n}$
represents the anisotropy of the superfluid density.

In a layered Q2D system, the renormalized polarization functions
$\tilde{Q}_{\mu \nu}$ are given by Eq.\ (\ref{tilde_Q}) with the
appropriate form of the Coulomb interaction~\cite{Das Sarma}:
\begin{eqnarray}
  V(\bm q)=\frac{2\pi d}{|\bm q_\||}
  \frac{\sinh(|\bm q_\||d)}{\cosh(|\bm q_\||d)-\cos(q_zd)},
\label{V_q}
\end{eqnarray}
where $\bm q_\|=(q_x,q_y)$. Equation\ (\ref{V_q}) reduces to the 3D
expression for the Coulomb potential $V_{\rm 3D}(\bm q)$ at small
momenta $|\bm q|d\ll 1$ and to the 2D Fourier transform of the
Coulomb potential $V_{\rm 2D}(\bm q_\|)$ in the limit $|\bm
q_\||d\gg 1$ as follows:
\begin{equation}
  V_{\rm 3D}(\bm q)=\frac{4\pi}{\bm q^2},  \quad
  V_{\rm 2D}(\bm q_\|)=\frac{2\pi d}{|\bm q_\||}.
\label{V_q-23D}
\end{equation}
In the long-wavelength limit ($|\bm q|d\ll 1$) considered in the rest
of our paper, the appropriate form of the Coulomb potential is $V_{\rm
3D}(\bm q)$.

\subsection{Linear response}
\label{Sec:linear-response}

By taking a variation of Eq.\ (\ref{S(A)}) with respect to $A^\mu$,
we obtain the gauge-invariant electromagnetic response of the
system,
\begin{equation}
  j_\mu(q) = K_{\mu\nu}(q) A^\nu(q).
\label{j_mu3D}
\end{equation}
Notice that the identities (\ref{qK}) ensure that the current
(\ref{j_mu3D}) satisfies the continuity equation $q^\mu j_\mu=0$.

To obtain physical results, we perform an analytical continuation
from the Matsubara frequency to the real frequency
$i\Omega_n\rightarrow \omega+i\delta$ \cite{Euclidean}.  After the
continuation, the vector $q$ becomes $q^\mu=(\omega,\bm q)$, and
Eq.\ (\ref{j_mu3D}) gives a causal response to the external
electromagnetic field $A^\mu=(A_0,\bm A)$ at a finite temperature
$T$.  The kernel $K_{\mu\nu}(q)$ is defined via Eqs.\
(\ref{tilde_Q}) and (\ref{K}) in terms of the one-loop response
functions (\ref{Q00})-(\ref{Qj0a}).  Their analytical continuations
to the real frequency after summation over the fermionic Matsubara
frequencies are given in Appendix~\ref{App:Polarization}.

\section{Collective modes}
\label{Sec:Collective}

Under certain conditions, an infinitesimal external
electromagnetic field can induce a large current or density response
of the system, which indicates the existence of internal
collective excitations in the superconductor.  These resonances
occur when the denominator in Eq.~(\ref{K}) goes to zero.  Using
Eq.\ (\ref{tilde_Q}), the denominator can be written as
$\tilde{Q}_{\mu\nu}q^{\mu}q^{\nu}=\tilde R(\omega,\bm q)/[1-V(\bm
q)\,Q_{00}]$, where the function $\tilde R(\omega,\bm q)$ is
defined as
\begin{eqnarray}
  \tilde R(\omega,\bm q) &=& Q_{kl}q_kq_l + 2\omega Q_{0l}^{(s)}q_l
  + \omega^2Q_{00}
\nonumber\\
  &+& V(\bm q)\,(Q_{0k}^{(s)}Q_{l0}^{(s)} - Q_{00}Q_{kl})\,q_kq_l.
\label{R3D}
\end{eqnarray}
Notice that only the conventional tensor $Q_{l0}^{(s)}$ appears in
Eq.\ (\ref{R3D}).  The anomalous tensor $Q_{k0}^{(a)}$ does not
appear because it is transverse, as will be discussed in
Sec.~\ref{Sec:Q_k0-a}.  The dispersion relation for the collective
modes is determined by the equation $\tilde R(\omega,\bm q)=0$.

In the limit of small $|\bm q|$, we can simplify Eq.~(\ref{R3D}) by
keeping only the nonvanishing terms $Q_{00}$ and $Q_{kl}$ in the
first and second lines of Eq.\ (\ref{R3D}), whereas the terms
$Q_{k0}^{(s)}$ and $Q_{kl}^{(2)}$ vanish in this limit. Using
Eq.~(\ref{tensor-n}) for $Q_{kl}^{(1)}$, we find the dispersion
relation for the collective modes at $T=0$,
\begin{equation}
  \omega^2 = V(\bm q) \frac{e^2 n_{0}}{m_e d}
  \left( \bm q_{\|}^2 + \frac{t_c^2m_e d^2}{\varepsilon_F} \, q_z^2
  \right).
\label{coll3D}
\end{equation}
In the case $|\bm q|d\ll1$, it is appropriate to use the 3D limit
$V_{\rm 3D}(\bm q)$ from Eq.\ (\ref{V_q-23D}) for the Coulomb
interaction in Eq.\ (\ref{coll3D}) and, thus, we find the following
spectrum of collective modes \cite{Hirschfeld1993, Levin, Das
Sarma}:
\begin{equation}
  \omega^2 = \frac{\omega_{ab}^2\bm q_\|^2 + \omega_c^2q_z^2}
  {\bm q_\|^2 + q_z^2}
  = \omega_{ab}^2\,\frac{\bm q\cdot \tensor{n}\cdot\bm q}{\bm q^2}.
\label{coll3D2}
\end{equation}
Here, the tensor $\tensor{n}$ is defined in Eq.\ (\ref{tensor-n}),
and $\omega_{ab}$ and $\omega_c$ are the in-plane and out-of-plane
plasma frequencies,
\begin{equation}
  \omega_{ab}^2 = 4\pi\frac{e^2n_{0}}{m_e d}, \qquad
  \omega_c^2 = 2\pi\frac{e^2t_c^2dn_0}{\varepsilon_F}.
\label{plasma}
\end{equation}
As shown in Appendix~\ref{App:Collective}, the collective mode
(\ref{coll3D2}) corresponds to coupled plasma oscillations of the
internal electric potential and the superconducting phase $\Phi$.  The
plasma mode frequency (\ref{coll3D2}) depends on the ratio of the
in-plane $\bm q_\|$ and out-of-plane $q_z$
momenta~\cite{HwangDasSarma98}. Plasma oscillations are gapped in all
directions, but are strongly anisotropic due to the smallness of the
interlayer tunneling amplitude $t_c$.  The experimental values of the
plasma frequencies in $\rm Sr_2RuO_4$ are $\omega_{c}=0.32$ eV and
$\omega_{ab}=4.5$ eV \cite{Tokura96, Tokura01}.

If one formally sets $t_c=0$ and uses the 2D expression for the
Coulomb potential from Eq.\ (\ref{V_q-23D}), then Eq.\ (\ref{coll3D})
gives the plasmon mode dispersion for a single 2D layer
\begin{eqnarray}
\label{dispersion2D}
  \omega_{\bm q}=\sqrt{\frac{2\pi n_{0}e^2}{m_e}|\bm q_\||}.
\end{eqnarray}
However, this limit does not correspond to $\rm Sr_2RuO_4$.

At higher temperatures $T\rightarrow T_c$, there may be
long-wavelength oscillations with the acoustic spectrum.  These
oscillations are neutral and consist of supercurrent oscillations
compensated by oscillations of the normal current, as discovered by
Carlson and Goldman \cite{Goldman}.  However, at $T\ll T_c$ only the
plasma mode survives, because the normal density is exponentially
suppressed.

\section{CONVENTIONAL NONCHIRAL ELECTROMAGNETIC RESPONSE}
\label{Sec:Conventional}

\subsection{Transverse and longitudinal tensors}
\label{Sec:Q_mu_nu}

First, we describe general properties of the tensors $Q_{kl}$ and
$Q_{k0}^{(s)}$.  It is convenient to separate the symmetric tensor
$Q_{kl}$ into the longitudinal $Q_{kl}^\|$ and transverse
$Q_{kl}^\perp$ parts defined by the following relations
\begin{eqnarray}
  & Q_{kl} = Q_{kl}^\| + Q_{kl}^\perp, &
\label{Q|+perp} \\
  & q_kQ_{kl}^\|=q_kQ_{kl}, \quad Q_{kl}^\|q_l=Q_{kl}q_l, &
\label{eq:Q|} \\
  & q_kQ_{kl}^\perp=Q_{kl}^\perp q_l=0. &
\label{eq:Qperp}
\end{eqnarray}
Equations (\ref{eq:Q|})--(\ref{eq:Qperp}) have the following general
solution
\begin{equation}
   Q_{kl}^\|= \frac{Q_{kr}q_r q_sQ_{sl}}{q_uQ_{uv}q_v}, \quad
   Q_{kl}^\perp=Q_{kl} - Q_{kl}^\|.
\label{Q_||_perp}
\end{equation}
In the dynamic limit, $\omega\neq0$ and $\bm q\to0$, which is relevant
for optical measurements, the paramagnetic term $Q_{kl}^{(2)}$,
defined in Eq.\ (\ref{Qij_2}), vanishes, and $Q_{kl}$ is given by the
diamagnetic term~(\ref{tensor-n}).  Then, the transverse and
longitudinal parts of the tensor $Q_{kl}$ are
\begin{eqnarray}
  Q^{\perp}_{kl} &=& \frac{\omega_{ab}^2}{4\pi}
  \left(\tensor{n}-\frac{(\tensor{n}\cdot\bm q)(\bm q\cdot\tensor{n})}
  {\bm q\cdot\tensor{n}\cdot\bm q}\right)_{kl},
\label{Qperp} \\
  Q^{\|}_{kl} &=& \frac{\omega_{ab}^2}{4\pi}
  \left(\frac{(\tensor{n}\cdot\bm q)(\bm q \cdot\tensor{n})}
  {\bm q\cdot\tensor{n}\cdot\bm q}\right)_{kl}.
\label{Q|}
\end{eqnarray}

The conventional current-density correlation function
$Q_{k0}^{(s)}$, given by Eq.\ (\ref{Qj0_1}), is an odd function of
$\bm q$ and $\omega$.  The term $Q_{k0}^{(s)}$ satisfies the
following identity for small $\bm q$:
\begin{equation}
  Q_{0k}^{(s)} = Q_{kl}^\|q_l \, \frac{q_uQ_{u0}}{q_rQ_{rs}^\|q_s}.
\label{Q_0k-Q_kl}
\end{equation}
To prove Eq.\ (\ref{Q_0k-Q_kl}), we observe that the tensor
structure of $Q_{kl}=\int v_kv_l\ldots$ and $Q_{k0}^{(s)}=\int
v_k\ldots$ in Eqs.\ (\ref{Qij13D}), (\ref{Qij_2}), and (\ref{Qj0_1})
is determined by the electron velocities $\bm v=\partial \xi_{\bm
p}/\partial\bm p$. Since the term $Q_{k0}^{(s)}$ vanishes at $\bm
q=0$, the leading-order expansion of the integrand is proportional
to $v_kv_lq_l$, i.e.,\ $Q_{k0}^{(s)}=\int v_kv_lq_l\ldots$.  Thus,
in the long-wavelength limit, we have $Q_{k0}^{(s)}\propto
Q_{kl}q_l$, which leads to Eq.\ (\ref{Q_0k-Q_kl}).  Equation
(\ref{Qj0_1}) gives the following explicit expression for
$Q^{(s)}_{k0}$ for small $\bm q$ and finite $\omega$ at $T=0$:
\begin{eqnarray}
  && Q^{(s)}_{k0} = -\frac{\omega_{ab}^2}{4\pi}
  \frac{\omega(\tensor{n}\cdot\bm q)_k}{\Delta_0^2}
  \int_1^{\infty}\frac{dx}{x^2\sqrt{x^2-1}}
\label{Q_0k-s} \\
  && \times \frac{1}{2x+(\omega+i\delta)/\Delta_0}
  \, \frac{1}{2x-(\omega+i\delta)/\Delta_0}.
\nonumber
\end{eqnarray}
We see that $Q^{(s)}_{k0}$ is proportional to the tensor $\tensor{n}$
multiplied by $\bm q$, which is consistent with Eq.\
(\ref{Q_0k-Q_kl}).

\subsection{Conventional nonchiral electromagnetic response}
\label{Sec:conventional}

We now show that if we omit the anomalous chiral term (\ref{Qj0a})
in Eqs.\ (\ref{tilde_Q}) and (\ref{K}), we recover the conventional
response of a superconductor to the electromagnetic field.  Using
Eqs.\ (\ref{tilde_Q}) and (\ref{Q|+perp}) and the identity
(\ref{Q_0k-Q_kl}), we obtain the following expression for the
space-time components of the conventional response kernel
$K_{\mu\nu}^{(c)}$ (\ref{K}):
\begin{eqnarray}
  & K_{kl}^{(c)} = Q_{kl}^\perp-\omega^2 \tilde\kappa_{kl}, \quad
  K_{00}^{(c)} = - q_k \tilde\kappa_{kl} q_l, &
\label{K_c} \\
  & K_{l0}^{(c)} = K_{0l}^{(c)}=\omega q_k \tilde\kappa_{kl}. &
\nonumber
\end{eqnarray}
In the long-wavelength limit assumed here, the transverse tensor
$Q_{kl}^\perp$ is given by Eq.~(\ref{Qperp}), and the longitudinal
tensor $\tilde\kappa_{kl}$ is defined as follows:
\begin{equation}
  \tilde\kappa_{kl}=
  \frac{Q_{k0}^{(s)}Q_{0l}^{(s)} - Q_{kl}^\|Q_{00}}{\tilde R},
\label{kappa-tilde}
\end{equation}
with the function $\tilde R(\omega,\bm q)$ given by Eq.~(\ref{R3D}).
By using Eq.~(\ref{K_c}), the charge and current responses
(\ref{j_mu3D}) can be written in the following form:
\begin{eqnarray}
  \delta \rho &=& -i\bm q\cdot\bm P,
\label{charge} \\
  \bm j &=& - \, \tensor{Q}^\perp \cdot\bm A - i\omega\bm P,
\label{current} \\
    \bm P &=& \tensor{\tilde\kappa}\cdot\bm E^{\rm ext}.
\label{polarization}
\end{eqnarray}
Here $\delta \rho$ and $\bm j$ are the induced charge and current
densities, and $\bm P$ is the polarization vector. In
Eq.~(\ref{current}) we use the shorthand notation $ [\, \tensor
Q^{\perp} \, ]_{kl} \equiv Q^{\perp}_{kl} $~[see Eq.~(\ref{Qperp})].

Equation (\ref{polarization}) expresses $\bm P$ in terms of the
\emph{external} electric field $\bm E^{\rm ext} =-i\bm
qA_0+i\omega\bm A$~(We remind the reader that the electromagnetic
potential $A^\mu$ in our calculations represents the external
field.) $\bm E^{\rm ext}$ is connected to the \emph{total} electric
field $\bm E^{\rm tot}$, which includes the field created by other
electrons in the system, by the standard relation
\begin{equation}
  \bm E^{\rm ext} = \bm E^{\rm tot} + 4\pi\bm P
  = \tensor{\varepsilon}\cdot \bm E^{\rm tot},
\label{DandE}
\end{equation}
where $\tensor{\varepsilon}$ is the tensor of dielectric
permeability. If we expressed the polarization vector $\bm P$ in
Eq.\ (\ref{polarization}) as a linear function of the total electric
field $\bm E^{\rm tot}$, then the corresponding proportionality
tensor would be the dielectric susceptibility tensor $\kappa_{kl}$
\cite{Landau}.  However, because Eq.\ (\ref{polarization}) expresses
$\bm P$ as a function of the external field $\bm E^{\rm ext}$, the
corresponding tensor $\tilde\kappa_{kl}$ is the RPA-renormalized
dielectric susceptibility tensor, which includes the effect of
charge screening via the RPA diagrams shown in Fig.\ \ref{fig_diag}.

The tensor $\tilde\kappa_{kl}$ in Eq.\ (\ref{kappa-tilde}) can be
simplified in the limit of small $\bm q$.  In this case, the terms
involving $Q_{k0}^{(s)}$ in the numerator and denominator of
Eq.~(\ref{kappa-tilde}) can be neglected relative to the other terms
because $Q_{k0}^{(s)}$ vanishes at $\bm q\to0$.  Using this
approximation and Eq.~(\ref{R3D}), we rewrite
Eq.~(\ref{kappa-tilde}) in terms of the longitudinal tensor
$Q_{kl}^\|$ (\ref{Q|}) as follows:
\begin{eqnarray}
  \tensor{\tilde\kappa} &=& -\frac{\alpha(\omega,\bm q)}{\omega_{ab}^2}
  \,\tensor{Q}^\|,
\label{kappa'}  \\
  \alpha(\omega, \bm q)&=&\frac{\omega_{ab}^2 Q_{00}}
  {\bm q\cdot\tensor{Q}\cdot\bm q + \omega^2Q_{00} -
  V(\bm q)\,Q_{00}\,\bm q\cdot\tensor{Q}\cdot\bm q}.
  \label{alpha}
\end{eqnarray}
Here, the dimensionless function $\alpha(\omega,\bm q)$ describes
screening of charge in the static case and plasma oscillations in
the dynamic case.

Indeed, in the static limit $\omega=0$, the function $Q_{00}$, given
by Eq.~(\ref{Q00_1}), is proportional to the density of states
\begin{equation}
  Q_{00}(\omega=0) = - 2e^2 N_0/d, \quad N_0=m_e/2\pi,
\label{N_0}
\end{equation}
where $N_0$ is the 2D density of states per spin.  Then
Eq.~(\ref{alpha}) for $\omega=0$ reads
\begin{equation}
  \alpha^{(s)}(\omega=0, \bm q) = -\frac{1}{(\bm q\cdot\tensor{n}\cdot\bm q)}
  \frac{q_{{TF}}^2}{1+q_{\rm TF}^2/\bm q^2},
\label{alpha-static}
\end{equation}
where $q_{\rm TF}$ is the inverse Thomas-Fermi screening length
\begin{equation}
\label{ThomasF}
  q_{\rm TF}^2=8\pi e^2 N_0/d.
\end{equation}
Thus, Eq.\ (\ref{alpha-static}) describes the electrostatic
screening of charge in an anisotropic conductor.

In the dynamic case $\omega\gg v_F|\bm q|$, we neglect the first term
in the denominator of Eq.~(\ref{alpha}) and obtain
\begin{equation}
  \alpha^{(d)}(\omega,\bm q)= \frac{\omega_{ab}^2}
  {\omega^2-[\omega_{ab}^2 (\bm q\cdot\tensor{n}\cdot\bm q)/\bm q^2]}.
\label{alpha-dynamic}
\end{equation}
Equation (\ref{alpha-dynamic}) exhibits a resonance, signifying a
divergence of the charge response, when the frequency $\omega$
approaches to the plasma frequency defined in Eq.\ (\ref{coll3D2}).

Equations (\ref{kappa-tilde})--(\ref{polarization}) coincide with
the results of Ref.~\cite {UFN} [Cf. with Eqs.\ (28) and (29) of
Ref.~\cite {UFN}, which were obtained using an exact solution for
vertex functions]. Note that the transverse part of the current
response in Eq.\ (\ref{current}) is not affected by the collective
modes \cite{Schrieffer}, whereas the charge response Eq.\
(\ref{charge}) is strongly affected by the collective dynamics of
the superconducting phase and the Coulomb interaction.  Equations
(\ref{charge}) and (\ref{current}) satisfy the continuity equation
$\omega\,\delta \rho=\bm q\cdot\bm j$ and are invariant with respect
to gauge transformations of the electromagnetic field.  In
anisotropic superconductors, it is not practical to separate $\bm A$
and $\bm E^{\rm ext}$ into longitudinal and transverse components,
because such separation does not diagonalize the response
equations~(\ref{charge}) and (\ref{current}), unlike in the
isotropic case.

\section{Anomalous chiral electromagnetic response}
\label{Sec:Anomalous}

\subsection{Anomalous current-density correlation function}
\label{Sec:Q_k0-a}

The chiral anomalous current-density correlation function
$Q_{k0}^{(a)}$ is given by Eq.\ (\ref{Qchiral}).  By separating the
factors $q_x$ and $q_y$, it can be written as
\begin{equation}
  Q^{(a)}_{k0} = i\Theta
  \left(\begin{array}{c}
  q_y \\ -q_x \\ 0 \\
  \end{array}\right)_k
  = i\Theta \, \eta_{kl}q_l
  = i\Theta \, \check{q}_k,
\label{Q_0k-perp}
\end{equation}
where $\eta_{xy}=-\eta_{yx}=1$ is the 2D antisymmetric tensor, and
$\check{q}_k$ is the $k$th component of the vector
\begin{equation}
  \check{\bm q}=\hat {\bm z}\times\bm q.
\label{q_perp}
\end{equation}
The anomalous term $Q_{k0}^{(a)}$ (\ref{Q_0k-perp}) is transverse:
$q_kQ_{k0}^{(a)}=0$, unlike the conventional term $Q_{k0}^{(s)}$ in
Eqs.\ (\ref{Q_0k-Q_kl}) and (\ref{Q_0k-s}).

\begin{figure} \centering
\includegraphics[width=0.9\linewidth]{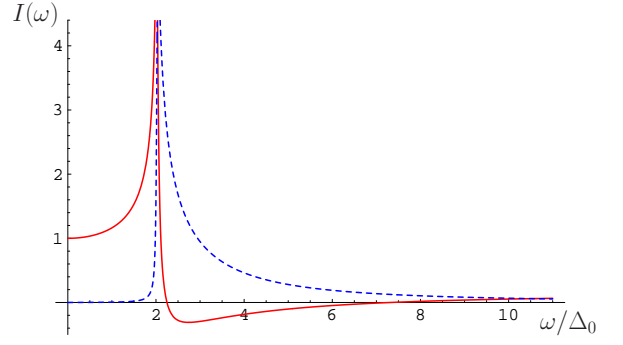}
  \caption{(color online).  Frequency dependence of the function
  $I(\omega)$ given by Eq.\ (\ref{I}).  This function determines the
    frequency dependence of the ac Hall conductivity
  (\ref{sigma_xy-w}) of a chiral $p_x+ip_y$ superconductor in the
  dynamic limit.  The solid (red) and dashed (blue) lines show the
  real and imaginary parts of $I(\omega)$.}
\label{figI1}
\end{figure}

In the limit $\bm q\to0$, we obtain the following expression for the
function $\Theta(\omega)$ from Eq.\ (\ref{Qchiral}):
\begin{eqnarray}
  \Theta(\omega) &=& \frac{e^2 \Delta_x\Delta_y}{m_e}
  \int{\frac{d^3 p}{(2\pi)^3}} \frac{p_x^2+p_y^2}{4E^2}\,[1-2f(E)]
\nonumber \\
  &\times& \left(\frac{1}{2E+\omega+i\delta}+\frac{1}{2E-\omega-i\delta}
  \right),
\label{Theta}
\end{eqnarray}
which was found before in Ref.~\cite{Yakovenko07}. Changing the
variable of integration to $x=E/\Delta_0$ and taking into account
that the integral converges near the Fermi surface, we express Eq.\
(\ref{Theta}) in the following form at $T=0$:
\begin{eqnarray}
  \Theta(\omega) &=& s_{xy}\,\frac{e^2}{2h d}\,I(\omega),
  \quad\quad I(0)=1,
\label{Theta'} \\
  I(\omega) &=& \int\limits_1^\infty \frac{dx}{\sqrt{x^2-1}} \,
  \frac{1}{x^2 - [(\omega+i\delta)/2\Delta_0]^2}.
\label{I}
\end{eqnarray}
We restored the dimensional constants in Eq.\ (\ref{Theta'}).  The
sign of $\Theta$ is determined by the chirality $s_{xy}$ of the
superconducting condensate.  The function $I(\omega)$ has the
following asymptotic behavior \cite{I(omega)}:
\begin{equation}
  I(\omega) \approx \left\{
  \begin{array}{ll} \displaystyle
    -4\left(\frac{\Delta_0}{\omega}\right)^2
    \ln\left(\frac{\omega}{\Delta_0}\right)
    + 2\pi i \left(\frac{\Delta_0}{\omega}\right)^2,
    & \omega\gg\Delta_0
  \\ \displaystyle
    1, & \omega\ll\Delta_0.
  \end{array}  \right.
\label{asymptotes}
\end{equation}
The crossover between the two limiting cases occurs at $\omega\sim
2\Delta_0$, where the photon energy is equal to the binding energy
between two electrons in a Cooper pair.  The real and imaginary parts
of $I(\omega)$ are shown in Fig.~\ref{figI1}.  The imaginary part is
zero for $\omega<2\Delta_0$ and diverges as
$1/\sqrt{\omega-2\Delta_0}$ when $\omega$ approaches $2\Delta_0$ from
above.  The real part is approximately constant at low frequencies and
diverges as $1/\sqrt{2\Delta_0-\omega}$ when $\omega$ approaches
$2\Delta_0$ from below.  Both real and imaginary parts go to zero at
$\omega\to\infty$.

\subsection{Anomalous chiral electromagnetic response}
\label{Sec:Anomalous-response}

Now we collect the terms that contain the chiral anomalous
correlator $Q_{k0}^{(a)}$ in the kernel $K_{\mu\nu}$ (\ref{K}).
First, we notice that $Q_{k0}^{(a)}$ does not appear in the
denominator $\tilde Q_{\alpha\beta}q^\alpha q^\beta$ of Eq.\
(\ref{K}), because the denominator is longitudinal, whereas
$Q_{k0}^{(a)}$ is transverse. Then, Eq.~(\ref{K}) contains only the
linear and quadratic terms in $Q_{k0}^{(a)}$.  The quadratic in
$Q_{k0}^{(a)}$ term may be called the ``double-anomalous.'' Although
this term has an anomalous origin, it is not chiral, i.e.,\ it does
not change sign under the time-reversal operation or when the
chirality $s_{xy}$ of the order parameter changes sign.  This term
turns out to be small in the relativistic parameter $(v_F/c)^2$ and
is not particularly important. The discussion of this term is
deferred to Appendix \ref{Sec:alt-action}.

In the rest of the paper, we concentrate on the linear in
$Q_{k0}^{(a)}$ terms in the response. Using Eq.~(\ref{K}), one
obtains the following expression for the space-time components of
the anomalous chiral kernel $K_{\mu\nu}^{(a)}$:
\begin{eqnarray}
  && K_{kl}^{(a)} =  i\omega \Theta
  \frac{\check q_k W_l-W_k\check q_l}{\tilde R},
  \quad  K_{00}^{(a)} = 0,
\label{K_a} \\
  && K_{l0}^{(a)}(q) = - K_{0l}^{(a)}(q) = i\Theta \check q_l
  \frac{q_k W_k}{\tilde R}.
\nonumber
\end{eqnarray}
Here, $\tilde R$, $\Theta$, and $\check q_k$ are given by Eqs.\
(\ref{R3D}), (\ref{Q_0k-perp}), and (\ref{q_perp}), and the vector
$\bm W$ is defined as
\begin{equation}
  W_k = Q_{kl}^\|q_l+Q_{k0}^{(s)}\omega.
\label{W}
\end{equation}
By using Eq.\ (\ref{K_a}), the anomalous charge and current
responses (\ref{j_mu3D}) can be written in the following form:
\begin{eqnarray}
  \delta \tilde{\rho}^{(a)} &=& -i\bm q \cdot \bm P^{(a)},
\label{charge-a} \\
  \bm j^{(a)} &=& i\bm q \times \bm M^{(a)} - i\omega\bm P^{(a)},
\label{current-a}
\end{eqnarray}
where $\bm {M}^{(a)}$ and $\bm {P}^{(a)}$ can be identified as the
chiral magnetization and polarization~\cite{White-Geballe, Landau}
and are given by
\begin{eqnarray}
  \bm P^{(a)} &=& - i \Theta\bm W
  \frac{[\bm q\times\bm E^{\rm ext}]\cdot\hat {\bm z}}{\omega \tilde R}
  =- i\Theta \frac{\bm W}{\tilde R} B^{\rm ext}_z,
\label{polarization-a} \\
  \bm M^{(a)} &=& i\hat {\bm z}\,\Theta
  \frac{(\bm W\cdot\bm E^{\rm ext})}{\tilde R}.
\label{magnetization-a}
\end{eqnarray}
Here, $B^{\rm ext}_z$ is the $z$ component of the external magnetic
field $\bm B^{\rm ext}$, which is related to $\bm E^{\rm ext}$ via
Maxwell's equation $\bm q\times\bm E^{\rm ext}=\omega\bm B^{\rm
ext}$. Both $\bm M^{(a)}$ and $\bm P^{(a)}$ in Eqs.\
(\ref{polarization-a}) and (\ref{magnetization-a}) are proportional
to the chiral response function $\Theta$ (\ref{Theta'}) and, thus,
change sign when the chirality $s_{xy}$ of the order parameter
changes sign.  The peculiar feature of chiral superconductors is
that the electric polarization $\bm P^{(a)}$ (\ref{polarization-a})
is induced by the magnetic field $B_z$, whereas the magnetization
$\bm {M}^{(a)}$ (\ref{magnetization-a}) couples to the longitudinal
electric field. The first and second terms in Eq.\ (\ref{current-a})
give the transverse and longitudinal components of the anomalous
current, respectively.

In the limit of small $|\bm q|$, by using Eqs.\ (\ref{Q|}) and
(\ref{Q_0k-s}) and the condition $I(0)=1$, the vector $\bm W$ in
Eq.\ (\ref{W}) can be written as
\begin{equation}
  W_k=q_lQ_{lk}^\| + \omega Q^{(s)}_{0k}
  = I(\omega)\,\frac{\omega_{ab}^2}{4\pi}(\tensor{n}\cdot\bm q)_k,
\label{combination}
\end{equation}
where $I(\omega)$ is given by Eq.\ (\ref{I}).  In the same limit, only
the terms proportional to $Q_{00}$ survive in the function $\tilde R$
given by Eq.\ (\ref{R3D}).  Using Eq.~(\ref{Q00_1}) at $\bm q\to0$, we
find that
\begin{equation}
  Q_{00}(\omega)=-\frac{2 e^2 N_0}{d}\,I(\omega),
\label{Q_00-w}
\end{equation}
where $I(\omega)$ again is given by Eq.\ (\ref{I}). Then, with the
help of Eqs.~(\ref{combination}), (\ref{Q_00-w}), and (\ref{R3D}),
we find that
\begin{equation}
  \frac{\bm W}{\tilde R}=- \frac{\alpha(\omega,\bm q)\,d}{4e^2 m_e} \,
  (\tensor{n}\cdot\bm q),
\label{eqn:U-alpha}
\end{equation}
where the function $\alpha(\omega,\bm q)$ is given by Eq.\
(\ref{alpha}). Using Eq.~(\ref{eqn:U-alpha}), the expressions
(\ref{polarization-a}) and (\ref{magnetization-a}) for $\bm P^{(a)}$
and $\bm M^{(a)} $ can be written as
\begin{eqnarray}
  \bm P^{(a)} &=& i\Theta \, \frac{\alpha(\omega,\bm q)\,d}{4e^2 m_e}
  \, \frac{[\bm q\times\bm E^{\rm ext}]\cdot\hat{\bm z}}{\omega}
  \, (\tensor{n}\cdot\bm q)
\label{polarization-as} \\
  &=& i\Theta \, \frac{\alpha(\omega,\bm q)\,d}{4e^2 m_e} \, B^{\rm ext}_z
  \, (\tensor{n}\cdot\bm q),
\nonumber \\
  \bm M^{(a)} &=& -i\hat{\bm z} \, \Theta \,
  \frac{\alpha(\omega,\bm q)\,d}{4e^2 m_e} \,
  (\bm q\cdot\tensor{n}\cdot\bm E^{\rm ext}).
\label{magnetization-as}
\end{eqnarray}
The static and dynamic limits of Eqs.~(\ref{polarization-as}) and
(\ref{magnetization-as}) can be obtained by taking appropriate
limits (\ref{alpha-static}) and (\ref{alpha-dynamic}) for the
function $\alpha(\omega,\bm q)$.

\subsection{Discussion of the anomalous magnetization current}
\label{Sec:magnetization}

By comparing Eq.\ (\ref{magnetization-as}) for the anomalous
magnetization with the expression for the induced conventional charge
density from Eqs.~(\ref{charge}), (\ref{polarization}),
(\ref{kappa'}), and (\ref{Q|}), one can notice that $\bm M^{(a)}$ can
be expressed in terms of $\delta\rho$
\begin{equation}
  \bm M^{(a)} = -s_{xy}\,\hat{\bm z}\,\frac{I(\omega)}{4m_e}\,\delta\rho.
\label{M-rho}
\end{equation}
Then, the first term in Eq.\ (\ref{current-a}), the magnetization
current, can be written in real space as
\begin{equation}
  \bm j_M^{(a)}(\omega,\bm r) = s_{xy} \, \frac{I(\omega)}{4m_e}
  \, [\hat{\bm z} \times \bm\nabla \delta\rho(\omega,\bm r)].
\label{j-rho}
\end{equation}
At $\omega=0$, Eq.\ (\ref{j-rho}) reduces to expression (\ref{j_M})
for the Mermin-Muzikar current~\cite{Mermin80}, which was discussed
in Sec.~\ref{Sec:Intro}.  However, at high frequencies, the
anomalous magnetization current (\ref{j-rho}) is suppressed by the
function $I(\omega)\propto(\Delta_0/\omega)^2$~[see
Eq.~(\ref{asymptotes})].  The general reason for this suppression is
that physical manifestations of the low-energy Cooper pairing with
the angular momentum $L_z=\hbar$ should fade away at high
frequencies $\omega\gg\Delta_0$.  Indeed, the electron states at
high energies are essentially the same as in a normal metal, so the
properties of the system should approach those of a normal metal at
high frequencies.  The suppression of the anomalous magnetization
current at high frequencies is one of the reasons why the Kerr angle
is so small \cite{Mineev07}.  Equation (\ref{j-rho}) gives an
important generalization of the Mermin-Muzikar current to arbitrary
frequencies.

In order to calculate the magnetization current correctly, it is
very important to take into account the term $Q^{(s)}_{0k}$ in Eq.\
(\ref{combination}).  This term vanishes at $\omega=0$ and can be
neglected at low frequencies, as was done in Refs.\
\cite{Goryo98,Goryo99,Goryo00,Golub03,Stone04}.  However, at high
frequencies, Eq.~(\ref{Q_0k-s}) shows that $Q^{(s)}_{0k}\propto
-q_k/\omega$, and, in the leading order of approximation, the second
term of the sum in Eq.\ (\ref{combination}) cancels the first term.
The remaining difference gives the small factor
$I(\omega)\propto(\Delta_0/\omega)^2$. The term $Q_{00}$ is also
proportional to $I(\omega)$ in Eq.\ (\ref{Q_00-w}), so the factors
$I(\omega)$ in $\bm W$ and $\tilde R$ cancel out in Eq.\
(\ref{eqn:U-alpha}), but $\Theta(\omega)$ produces the factors
$I(\omega)$ in Eqs.\ (\ref{magnetization-as}), (\ref{M-rho}), and
(\ref{j-rho}).  If $Q^{(s)}_{0k}$ were neglected in Eq.\
(\ref{combination}), then $\bm W$ would have a constant value at
high frequencies, and one would incorrectly conclude that
Eq.~(\ref{j_M}) is valid for arbitrarily high frequencies. The
asymptotic cancellation of the two terms in the sum in Eq.\
(\ref{combination}) can also be shown using the Ward identities
\cite{UFN}.

\subsection{Discussion of the anomalous polarization current}
\label{Sec:polarization}

The second term $\bm j_P^{(a)}$ in Eq.\ (\ref{current-a}) is
determined by the chiral polarization $\bm P^{(a)}$.  Although $\bm
j_P^{(a)}$ is formally proportional to $\omega$, it contributes
equally to the total current even at low frequencies.  Indeed,
Eqs.~(\ref{polarization-a}) and (\ref{polarization-as}) show that
the chiral polarization diverges at low frequencies as $1/\omega$.
This divergence exactly cancels $\omega$ in Eq.~(\ref{current-a}),
which results in a finite contribution from the chiral polarization
to the current even at low frequencies.  Both magnetization and
polarization currents equally contribute to the Hall conductivity
tensor, which is discussed below in Eqs.\ (\ref{sigma_a}) and
(\ref{sigma_xy}). The anomalous polarization current $\bm j_P^{(a)}$
was often omitted in the previous papers in the low frequency limit.

Equations (\ref{charge-a}), (\ref{polarization-a}), and
(\ref{polarization-as}) show that the magnetic field $B_z$ induces an
anomalous electric charge, as mentioned in Sec.\ \ref{Sec:Intro}.  The
polarization current $\bm j_P^{(a)}$ in Eq.\ (\ref{current-a}) is
necessary to satisfy the continuity equation for the anomalous charge
in Eq.\ (\ref{charge-a}).

\subsection{Anomalous chiral effective action}
\label{Sec:anomalous-action}

Substituting the components (\ref{K_a}) of the anomalous tensor
$K_{\mu\nu}^{(a)}$ into Eq.\ (\ref{S(A)}), we obtain the chiral part
of the gauge-invariant effective action for the electromagnetic
field as
\begin{eqnarray}
  && \tilde S_{\rm eff}^{(a)} = -i \sum_q \Theta(q) \,
  \frac{\bm W(q)\cdot\bm E^{\rm ext}(q)}{\tilde R(q)}
  \,B^{\rm ext}_z(-q)
\nonumber \\
  && \approx i \frac{s_{xy}}{8hm_e} \sum_q \alpha(\omega,\bm q)\,I(\omega)\,
  [\bm q\cdot\tensor{n}\cdot\bm E^{\rm ext}(q)] \, B^{\rm ext}_z(-q).
\label{tilde_S_a}
\end{eqnarray}
Given that Eq.\ (\ref{K_a}) is written for the real frequency
$\omega$, we write Eq.~(\ref{tilde_S_a}) for the real frequency as
well.  The anomalous charge and current responses, given by
Eqs.~(\ref{charge-a})--(\ref{magnetization-a}), can be obtained by
taking the appropriate variations of Eq.\ (\ref{tilde_S_a}). The
causality of the response function should be properly addressed, as
discussed in Ref.~\cite{Negele}.

Unlike the conventional action for the electromagnetic field, the
anomalous action (\ref{tilde_S_a}) involves a product of the
electric and magnetic fields.  Thus, the anomalous action
(\ref{S_a}) breaks the TRS, because $\bm E\to\bm E$ and $\bm
B\to-\bm B$ upon the time-reversal operation.  Equation (\ref{S_a})
is manifestly gauge-invariant and is a replacement for the
Chern-Simons-type term (\ref{CS}) after integration out of the
superconducting phase $\Phi$. The calculation of this action is one
of the central results of our paper. A simplified alternative
derivation of the effective action is also given in Appendix
\ref{Sec:alt-action}.

Anomalous effective actions of the forms similar to
Eq.~(\ref{tilde_S_a}) were obtained for chiral superfluids in
Refs.~\cite{Goryo98,Goryo99,Golub03}.  Coupling between the electric
and magnetic fields was discussed in Ref.\ \cite{Ishikawa98}.
However, because these calculations were performed in the
low-frequency limit, they did not obtain the factor $I(\omega)$,
which suppresses the chiral effects at high frequencies, as
discussed in Sec.~\ref{Sec:magnetization}.  The factor $\tilde R$ in
the denominator of Eq.\ (\ref{tilde_S_a}) represents the collective
modes of the system.  References \cite{Goryo98,Goryo99,Golub03} did
not take into account the Coulomb interaction, which is appropriate
for electrically neutral superfluids, such as $^3$He.  Thus, the
collective modes in the denominator of the effective actions in
these papers were the acoustic modes of the superconducting phase,
which can be obtained by setting $V(\bm q)=0$ in Eq.~(\ref{R3D}).
However, in the case of a charged superconductor, the collective
modes are the gapped plasmons represented by the function $\tilde R$
in Eq.~(\ref{tilde_S_a})~(see also Ref.~\cite{Golub03}). One should
keep in mind that, because we have integrated out the internal
Coulomb potential, the electromagnetic field in Eq.\
(\ref{tilde_S_a}) is the external one. Depending on the physical
context, it may also be useful to study the chiral response with
respect to the total electromagnetic field, which includes both
external and internal fields.  This problem is discussed in Sec.
\ref{Sec:Total}.

\section{Conductivity tensor of a chiral superconductor}
\label{Sec:Total}

\subsection{Response to the external versus total electric field}
\label{Sec:Total-external}

In order to calculate the polar Kerr angle (see
Appendix~\ref{Sec:Kerr-Hall}) and other observable experimental
quantities, one needs to know the conductivity tensor
$\tensor{\sigma}$ of a chiral superconductor, which is defined by the
standard relation
\begin{eqnarray}
  \bm j=\tensor{\sigma}\cdot\bm E^{\rm tot}.
\label{sigma-definition}
\end{eqnarray}
Here, $\bm E^{\rm tot}$ is the total electric field inside the
superconductor.  In principle, the tensor $\tensor{\sigma}$ can be
extracted from Eqs.~(\ref{current}) and~(\ref{current-a}) for the
conventional and chiral currents.  However, the polarization and
magnetization vectors in these equations are expressed by Eqs.\
(\ref{polarization}), (\ref{polarization-a}), and
(\ref{magnetization-a}) in terms of the external electric field $\bm
E^{\rm ext}$.  These equations give the linear response relation in
the form
\begin{eqnarray}
  \bm j=\tensor{\tilde\sigma}\cdot\bm E^{\rm ext}
\label{tilde-sigma}
\end{eqnarray}
with a different tensor $\tensor{\tilde\sigma}$ \cite{Mahan}, which
includes the renormalization due to the RPA diagrams shown in Fig.\
\ref{fig_diag}.

The linear response relation in the form (\ref{tilde-sigma}) is
physically transparent, because it gives a direct response of the
system to the external perturbation.  However, in general, the
response of the system to the external field depends on the geometry
of the sample and the experimental apparatus.  Only in the idealized
case of an infinite uniform system can the problem be solved by the
Fourier transform.  To deal with this problem, the standard approach
in the electrodynamics of continuous media \cite{Landau} is to use
the constituency relation (\ref{sigma-definition}), which expresses
the response of the media to the total electromagnetic field in
terms of the conductivity tensor characterizing the material. The
constituency relations are substituted into Maxwell's equations as
the source terms, and then Maxwell's equations are solved with the
boundary conditions appropriate for the experimental setup. This is
how, for example, one can take into account the Meissner screening,
which was not included in the RPA-renormalized linear response given
by Eq.~(\ref{tilde-sigma}).

In order to obtain the proper constituency relations, we need to
transform the results of the paper to the form
(\ref{sigma-definition}).  This can easily be done by noticing that,
in Eq.~(\ref{L_el}), the internal (induced) electric potential
$\varphi$ appears next to the external potential $A_0$.  Thus, the
combined electric potential corresponds to the total electric field
$\bm E^{\rm tot}$ inside the superconductor.  Then,
Eq.~(\ref{action1}) gives the effective action for the total
electromagnetic field. By taking a variational derivative with
respect to the total electromagnetic field, one arrives at
Eq.~(\ref{sigma-definition}).  Formally, this means that we do not
integrate out the internal field $\varphi$ and do not perform the
RPA renormalization of the response kernels $Q_{\alpha\beta}$ in
Eq.\ (\ref{tilde_Q}).  Thus, the transition between
Eqs.~(\ref{tilde-sigma}) and (\ref{sigma-definition}) can be
accomplished by setting $V(\bm q)=0$ and replacing $\bm E^{\rm ext}$
with $\bm E^{\rm tot}$ in the electromagnetic response.  The Coulomb
potential $V(\bm q)$ appears explicitly only in the function $\tilde
R(\omega,\bm q)$ defined in Eq.\ (\ref{R3D}).  Therefore, one should
replace this function with the bare one \cite{acoustic},
\begin{eqnarray}
  R(\omega,\bm q) &=& Q_{kl}q_kq_l + 2\omega Q_{0k}^{(s)}q_k
  + \omega^2Q_{00}.
\label{R}
\end{eqnarray}

Using this prescription and Eqs.~(\ref{current})
and~(\ref{current-a}), one can easily obtain the conductivity tensor
\begin{eqnarray}
  \tensor{\sigma}=\tensor{\sigma}^{(c)}+\tensor{\sigma}^{(a)},
\label{Eq:sigma}
\end{eqnarray}
which consists of the conventional (nonchiral)
$\tensor{\sigma}^{(c)}$ and the anomalous (chiral)
$\tensor{\sigma}^{(a)}$ contributions, as discussed below.

\subsection{Conventional nonchiral conductivity tensor}
\label{Sec:Total-conventional}

Following the prescription of Sec.\ \ref{Sec:Total-external} and
using Eqs.~(\ref{kappa-tilde}), (\ref{current}) and
(\ref{polarization}), we obtain the conventional part of the
conductivity tensor as follows:
\begin{eqnarray}
  \tensor{\sigma}^{(c)}&=&\tensor{\sigma}^{(1)}+\tensor{\sigma}^{(2)},
\\
  \tensor{\sigma}^{(1)}_{kl}&=&-\frac{1}{i\omega}Q_{kl}^\perp,
\\
  \tensor{\sigma}^{(2)}_{kl}&=& - i\omega
  \frac{(Q_{k0}^{(s)}Q_{0l}^{(s)}-Q_{kl}^\|Q_{00} )}{R},
\label{sigma_perp}
\end{eqnarray}
where the function $R(\omega,\bm q)$ is given by Eq.\ (\ref{R}).

In the dynamic limit $\omega\neq 0$ and $\bm q\to 0$, the terms
$Q_{k0}^{(s)}$ in the numerator of Eq.\ (\ref{sigma_perp}) and the
terms proportional to $\bm q$ in Eq.\ (\ref{R}) vanish. As a result,
the expression for the conductivity tensor is simplified
\begin{equation}
  \tensor{\sigma}^{(1)}_{kl} = -\frac{1}{i\omega}Q_{kl}^\perp ,
  \qquad \tensor{\sigma}^{(2)}_{kl} = -\frac{1}{i\omega}Q_{kl}^\|.
\label{sigma12}
\end{equation}
From Eq.~(\ref{sigma12}), we find the total conductivity tensor in
the long wavelength limit as follows:
\begin{eqnarray}
  \tensor{\sigma}^{(c)}_{kl} = - \frac{Q_{kl}^{(1)}}{i\omega}
  = -\frac{1}{4\pi
  i\omega} \left(\begin{array}{ccc} \omega_{ab}^2 & 0 & 0 \\ 0 &
  \omega_{ab}^2 & 0 \\ 0 & 0 & \omega_c^2
  \end{array}\right),
\label{tensor-sigma}
\end{eqnarray}
where we used Eq.\ (\ref{tensor-n}) for $Q_{kl}^{(1)}$.

Combining Eq.\ (\ref{tensor-sigma}) with the standard formula for
the dielectric permeability tensor $\tensor{\epsilon}$,
\begin{equation}
  \tensor{\epsilon}=\tensor{1}+\frac{4\pi i}{\omega}\tensor{\sigma},
\label{epsilon}
\end{equation}
we obtain
\begin{eqnarray}
  &&  \tensor{\epsilon}(\omega)=
    \left(\begin{array}{ccc}
    1-\frac{\omega_{ab}^2}{\omega^2} & 0 & 0 \\
    0 & 1-\frac{\omega_{ab}^2}{\omega^2} & 0 \\
    0 & 0 & 1-\frac{\omega_c^2}{\omega^2}
  \end{array}\right).
\label{tensor-epsilon}
\end{eqnarray}
Equations (\ref{tensor-sigma}) and (\ref{tensor-epsilon}) represent
the standard Drude response of an anisotropic conductor.

\subsection{Anomalous chiral conductivity tensor}
\label{Sec:Total-anomalous}

The anomalous chiral response of the system to the total electric
field is obtained from
Eqs.~(\ref{charge-a})--(\ref{magnetization-a}) by replacing $\tilde
R\to R$, where the function $R(\omega,\bm q)$ is given by Eq.\
(\ref{R}).  From Eqs.\ (\ref{current-a})--(\ref{magnetization-a})
for the current response, we obtain the anomalous chiral
conductivity tensor
\begin{eqnarray}
  \tensor{\sigma}^{(a)}_{kl} = \Theta\,
  \frac{\check q_k W_l -W_k\check q_l}{R},
\label{sigma_a}
\end{eqnarray}
where $\check{\bm q}$ and $W_k$ are given by Eqs.\ (\ref{q_perp}) and
(\ref{W}).  Notice that the anomalous conductivity tensor is
antisymmetric $\sigma^{(a)}_{kl}(q)=-\sigma^{(a)}_{lk}(q)$ and
represents the intrinsic Hall conductivity of a chiral superconductor.

Using Eq.~(\ref{combination}), we can rewrite the anomalous
conductivity tensor (\ref{sigma_a}) as
\begin{eqnarray}
  && \tensor{\sigma}^{(a)} =
  \frac{\omega_{ab}^2\Theta(\omega)I(\omega)}{4\pi R(\omega,\bm q)}
  \left[\check{\bm q} \, (\bm q\cdot\tensor{n}) - (\tensor{n}\cdot\bm q) \,
  \check{\bm q} \right]
\label{sigma_a'} \\
  && = \frac{\omega_{ab}^2\Theta I}{4\pi R}
  \left( \begin{array}{ccc}
  0 & -(q_x^2+q_y^2) & -\zeta q_yq_z \\
  q_x^2+q_y^2 & 0 & \zeta q_xq_z \\
  \zeta q_yq_z & -\zeta q_xq_z & 0
  \end{array} \right),
\label{sigma_a-matrix}
\end{eqnarray}
where $\zeta=\omega_c^2/\omega_{ab}^2$ is a small parameter
representing anisotropy of the band dispersion in a Q2D metal.  The
tensor components $\sigma^{(a)}_{zx}$ and $\sigma^{(a)}_{zy}$ in Eq.\
(\ref{sigma_a-matrix}) are small, unless $q_z^2\gg \zeta^2
(q_x^2+q_y^2)$.

Now we concentrate on the $\sigma^{(a)}_{xy}$ component of the
anomalous conductivity tensor~(\ref{sigma_a-matrix})
\begin{equation}
  \sigma^{(a)}_{xy} =- \Theta(\omega) \,
  \frac{\omega_{ab}^2\,I(\omega)}{4\pi R(\omega,\bm q)}\,q_\|^2 ,
  \quad q_\|^2=q_x^2+q_y^2.
\label{sigma_xy}
\end{equation}
It was argued in Refs.\ \cite{Yakovenko07,Mineev07} that the Hall
conductivity of a chiral superconductor can be obtained by omitting
the last term in the brackets of Eq.~(\ref{j_xy'}), which gives
$\sigma^{(a)}_{xy}=\Theta(\omega)$.  However, Eq.~(\ref{sigma_xy})
shows that the systematic integration out of the superconducting
phase $\Phi$ produces a different expression for
$\sigma^{(a)}_{xy}$, which differs from $\Theta(\omega)$ by the
additional factor proportional to $q_\|^2$.  This factor involves
the in-plane wave vector of the electromagnetic field and, in the
appropriate limit, makes $\sigma^{(a)}_{xy}$ vanish when
$q_\|^2\to0$.  This fact reflects the cancellation of the Hall
effect \cite{Kallin} for a uniform (in the plane) system discussed
in Sec.\ \ref{Sec:Intro}.  Equation (\ref{sigma_xy}) is the central
result of our paper and will be used in Sec.~\ref{Sec:experiment} to
estimate the observable Kerr effect. At low frequencies
$\omega\ll\Delta_0$, where $I\approx1$, Eq.~(\ref{sigma_xy}) agrees
with the corresponding results of Refs.~\cite{Golub03, Goryo98}.
However, at higher frequencies $\omega\geq\Delta_0$, the function
$I(\omega)$ exhibits the nontrivial frequency dependence shown in
Fig.~\ref{figI1}.  This behavior originates from the tendency to
cancel between the current-current $Q_{kl}$ and current-density
$Q^{(s)}_{k0}$ polarization functions in Eq.~(\ref{combination}).
Thus, it is essential to take $Q^{(s)}_{k0}$ into account at high
frequencies~(see discussion in Sec.~\ref{Sec:magnetization}).

Let us discuss Eq.\ (\ref{sigma_xy}) in different limits.  First, we
consider the static limit $\omega=0$, set $q_z=0$, and then take
$\bm q_\|\to0$.  Using Eqs.\ (\ref{Theta'}) and (\ref{R}), we see
that $q_\|^2$ cancels out, and we find that
$\sigma^{(a)}_{xy}=e^2/2h d$, which is reminiscent of the quantum
Hall effect \cite{Volovik88,Mineev07,Yakovenko07}.  However, this
formally calculated value does not correspond to an observable dc
Hall effect. The static limit describes the system in thermodynamic
equilibrium, where an applied electric field causes an inhomogeneous
equilibrium redistribution of the electron density.  As a result of
the nonzero Cooper-pair angular momentum, the inhomogeneous electron
density produces the equilibrium magnetization current
Eq.~(\ref{j-rho}). However, the total magnetization current flowing
through any cross section of the sample, including the bulk and the
edges, is zero, because the current is solenoidal.  Therefore, the
total current measured by an ammeter is zero. Thus, the formally
calculated $\sigma^{(a)}_{xy}$ does not represent a measurable Hall
effect \cite{dcHall}.

The experimentally relevant limit is the dynamic limit with
$\omega\neq0$ and small $\bm q$.  Taking this limit in
Eq.~(\ref{sigma_xy}) and using Eqs.\ (\ref{R}) and (\ref{Q_00-w}), we
find
\begin{equation}
  \sigma^{(a)}_{xy} =
  \Theta(\omega)\,\frac{v_F^2q_\|^2}{2\omega^2},
\label{sigma_xy-w}
\end{equation}
where $v_F$ is the in-plane Fermi velocity.  Equation
(\ref{sigma_xy-w}) gives the ac Hall conductivity, where the frequency
dependence of $\Theta(\omega)$ is given by Eqs.\ (\ref{Theta'}) and
(\ref{I}).  Equation (\ref{sigma_xy-w}) differs from the expression
$\sigma^{(a)}_{xy}=\Theta(\omega)$ obtained in the previous papers
\cite{Yakovenko07,Mineev07} by the small factor
$v_F^2q_\|^2/2\omega^2$.

\subsection{Anomalous effective action and charge response}
\label{Sec:charge}

Equation (\ref{tilde_S_a}) gives the anomalous effective action as a
function of the external electromagnetic field. Transformation of
this action to the total field is straightforward by replacing
$\tilde R\to R$ \cite{acoustic},
\begin{eqnarray}
  && S_{\rm eff}^{(a)} = -i \sum_q \Theta(q) \,
  \frac{\bm W(q)\cdot\bm E^{\rm tot}(q)}{R(q)}
  \,B^{\rm tot}_z(-q)
\nonumber \\
  && \approx i \frac{v_F^2}{2} \sum_q \Theta(\omega) \,
  \frac{[\bm q\cdot\tensor{n}\cdot\bm E^{\rm tot}(q)] \,B^{\rm tot}_z(-q)}
  {\omega^2-v_F^2(\bm q\cdot\tensor{n}\cdot\bm q)/2}.
\label{S_a}
\end{eqnarray}
The difference between Eqs.\ (\ref{tilde_S_a}) and (\ref{S_a}) is
that the former involves the gapped plasmon modes represented by
$\tilde R$ (\ref{R3D}) in the denominator, whereas the latter
involves the acoustic modes of the superconducting phase represented
by $R$ (\ref{R}).  This difference occurs because the two effective
actions are written using the screened and unscreened electric
fields, whereas the magnetic field is not screened by the internal
Coulomb potential. For small $\bm q$ and high $\omega$, this
difference amounts to using $R\approx Q_{00}\omega^2$ vs $\tilde
R\approx Q_{00}[\omega^2-\omega_{ab}^2 (\bm q\cdot\tensor{n}\cdot\bm
q)/\bm q^2]$, where the last term represents the momentum-dependent
plasma frequency.  The effective action is also discussed in
Appendix \ref{Sec:alt-action}.

By taking a variational derivative of the action (\ref{S_a}) with
respect to the total field $A_0^{\rm tot}$, we find the anomalous
electric charge induced by the magnetic field \cite{charge},
\begin{equation}
  \delta\rho^{(a)} = - \frac{\omega_{ab}^2}{4\pi} \,
  \frac{\Theta I}{R} \, (\bm q \cdot\tensor{n}\cdot \bm q) \, B_z^{\rm tot}.
\label{charge-response}
\end{equation}
In the static limit, Eq.~(\ref{charge-response}) reads
\begin{eqnarray}
  \delta\rho^{(a)} = - s_{xy} \, \frac{e^2}{2hdc} \, B_z^{\rm tot}.
\label{charge-response-static}
\end{eqnarray}
Equation (\ref{charge-response-static}) was derived earlier in
Refs.~\cite{Goryo00,Stone04} and was mentioned in
Sec.~\ref{Sec:Intro}.  It is reminiscent of the \u{S}treda formula
for the quantum Hall effect~\cite{Streda}.  However, at high
frequencies, Eq.~(\ref{charge-response}) shows that the induced
electric charge is significantly reduced relative to the static
limit (\ref{charge-response-static})
\begin{equation}
  \delta\rho^{(a)} = s_{xy} \, \frac{e^2}{2hdc} \, I(\omega) \,
  \frac{v_F^2(\bm q\cdot\tensor{n}\cdot\bm q)}{2\omega^2} \,
  B_z^{\rm tot}.
\label{charge-response-dynamic}
\end{equation}

It is instructive to calculate how much electric charge would be
induced by the static magnetic field of one superconducting vortex.
If we take an integral $\int dx\,dy$ of
Eq.~(\ref{charge-response-static}), the left-hand side gives us the
total induced electric charge, and the right-hand side gives the total
magnetic flux.  Taking the latter to be one superconducting flux
quantum $\phi_0=hc/2e$, we find that the total induced electric charge
in a vortex $\delta Q_{\rm vortex}$ is
\begin{equation}
  \delta Q_{\rm vortex} = - s_{xy} \, \frac{e}{4}
\label{e/4}
\end{equation}
per layer, i.e.,\ the superconducting vortex has a fractional
electric charge.  Equation (\ref{e/4}) was obtained by Goryo
\cite{Goryo00}, who also pointed out that a vortex has a fractional
angular momentum.  However, the anomalous induced charge (\ref{e/4})
would be screened by the conventional screening mechanism
\cite{charge}, and its experimental measurement may be challenging.
It should be emphasized that the charge density
(\ref{charge-response-static}) is not concentrated in the vortex
core at the coherence length, but extends to the London penetration
length, where the magnetic field is present.  A possible experiment
for detection of the anomalous induced electric charge is discussed
in Sec.\ \ref{Sec:AFM}.

\section{Experimental implications}
\label{Sec:experiment}

\subsection{Polar Kerr effect experiment}
\label{Sec:Estimates}

In this section, we apply the derived theoretical results to the
interpretation of the Kerr effect measurements \cite{Xia06}. We use
the relationship between the Hall conductivity $\sigma_{xy}$ and the
Kerr angle $\theta_K$ as presented in Appendix~\ref{Sec:Kerr-Hall}.

First, we briefly discuss the estimates of the polar Kerr angle in
the previous literature~\cite{Yakovenko07,Mineev07}, where the
formula $\sigma^{(a)}_{xy}=\Theta(\omega)$ was used without the
additional factor appearing in Eq.\ (\ref{sigma_xy-w}).  To estimate
the Kerr angle, Ref.~\cite{Yakovenko07} implicitly assumed that the
refraction coefficient $n$ is real by using the value $n(n^2-1)=3$
quoted in Ref.~\cite{Xia06} and expressed $\theta_K$ in terms of
$\sigma_{xy}''$ via Eq.\ (\ref{theta_K}).  The theoretical estimate
given in Ref.\ \cite{Yakovenko07} is $\theta_K\approx 230$ nrad,
which is of the same order of magnitude as the experimental value of
65 nrad \cite{Xia06}. In Ref.~\cite{Mineev07}, a more detailed
estimate was presented, using Eq.~(\ref{n}) for the refraction
coefficient $n$ and discussing different limits $\omega>\omega_p$
and $\omega<\omega_p$. However, the factor $\epsilon_\infty$ was
overlooked in Ref.~\cite{Mineev07} as well as the presence of the
real part $\sigma_{xy}'(\omega)$, which gives the primary
contribution to $\theta_K$ for $\omega<\omega_p$~( see
Appendix~\ref{Sec:Kerr-Hall}). Overall, the numerical estimate of
$\theta_K$ given in Ref.~\cite{Mineev07} is of the same order of
magnitude as in Ref.\ \cite{Yakovenko07}.

However, as discussed in Sec.\ \ref{Sec:Intro},
Refs.~\cite{Yakovenko07} and \cite{Mineev07} did not take into
account the self-consistent dynamics of the superconducting phase
$\Phi$ and, thus, missed the additional factor in
Eq.~(\ref{sigma_xy-w}), which makes the Hall conductivity dependent
on the wave vector $q_\|$. In this case, strictly speaking, one
cannot use the equations for $\theta_K$ presented in
Appendix~\ref{Sec:Kerr-Hall}, because they were derived for the
normally-incident infinite plane wave with $q_\|=0$.  However, as
shown in Fig.~\ref{fig:setup}, the experiment \cite{Xia06} was
performed with a tightly focused Gaussian laser beam, which has
nonzero Fourier components with $q_\|\neq0$. Solving the boundary
value problem for a reflection of a finite-size Gaussian beam from a
chiral superconductor and determining the Kerr angle for
polarization rotation is a complicated problem, that is beyond the
scope of this paper.  Nevertheless, to make a crude estimate of the
Kerr angle, we can use the equations from
Appendix~\ref{Sec:Kerr-Hall} and the Hall conductivity given by Eq.\
(\ref{sigma_xy-w}) with the replacement $q_\|\to1/l$, where $l$ is
the typical transverse size of the Gaussian beam~( see
Fig.~\ref{fig:setup}).  Taking into account that $\omega=c|\bm q|$,
where the wave vector $|\bm q|=2\pi/\lambda$ is related to the
wavelength of light $\lambda$, we can roughly estimate the Hall
conductivity in Eq.\ (\ref{sigma_xy-w}) as
\begin{equation}
  \sigma_{xy} \sim
  \Theta(\omega)\left(\frac{v_F}{c}\right)^2
  \left(\frac{\lambda}{l}\right)^2.
\label{sigma_xy-crude}
\end{equation}
Using the values $v_F=5.5\times10^4$ m/s for the $\gamma$ sheet of
the Fermi surface \cite{Mackenzie03}, $c=3\times10^8$ m/s,
$\lambda=1.55$ $\mu$m, $l\sim25$ $\mu$m \cite{Xia06}, and
$\Delta_0=0.8$ meV ~\cite{Sengupta02} in Eqs.~(\ref{theta_K-low})
and (\ref{sigma_xy-crude}), we estimate the Kerr angle to be
$\theta_K\sim 10^{-14}$~rad.  This estimate for $\theta_K$ is about
6 orders of magnitude smaller than the experimental value of $65$
nrad.  The strong suppression of the Kerr angle relative to the
previous estimate originates primarily from the small relativistic
factor $(v_F/c)^2$ in Eq.\ (\ref{sigma_xy-crude}).

In the rest of this section, we speculate about possible ways of
resolving the discrepancy between the theory and experiment. As
discussed in Sec.~\ref{Sec:Intro}, the general arguments of
Ref.~\cite{ReadGreen} show that the intrinsic Hall conductivity
should vanish for a spatially homogeneous uniform system.  One way
to break the translational symmetry is by taking into account the
finite size $l$ of the laser beam.  Then, inevitably, the Kerr angle
acquires a dependence on $l$ as shown in Eq.~(\ref{size}), which is
a very robust theoretical result.  This proportionality relation
should be checked experimentally.

Although the dependence on $l$ introduces a small factor, there may
be mechanisms for enhancement of the response of the system, which
may compensate for this additional smallness.  One possibility is a
resonance with the plasma modes. Equation (\ref{j-rho}) shows that
the magnetization current is produced by the gradients of electron
density.  Therefore, the problem of the chiral current calculation
reduces to how much electron charge is induced on the surface of a
crystal by the inhomogeneous laser beam.  The continuity of the
electric field lines in a Gaussian beam in vacuum requires that the
electric field components $E_z$ must be present at the periphery of
the beam, even though the electric field is nominally polarized
along $E_x$ at the center of the beam.  This is well known for the
fiber-optics modes, which have similarities with the Gaussian mode
in vacuum \cite{Meschede}. When the Gaussian beam hits the sample
(see Fig.~\ref{fig:setup}), the $E_z$ components induce electric
charges of opposite signs on the two sides of the beam in the $x$
direction, which induce the electric current $j_y\propto q_xE_z$
according to Eq.\ (\ref{j-rho}).  The induced electric current $j_y$
generates a magnetic field $B_x$ and a reflected electromagnetic
wave with the $E_y$ polarization.  Taking into account that
$E_z\propto(q_\|/|\bm q|)E_x$, we obtain the same result as in Eq.\
(\ref{sigma_xy-w}).  However, because the external electric field
$E_z$ directly couples to the induced electron charge, the response
of the system would be enhanced by the factor
$\alpha^{(d)}(\omega,\bm q)$ in Eq.~(\ref{alpha-dynamic}) when the
frequency of light $\omega$ is in resonance with one of the plasma
modes.  As discussed in Appendix \ref{Sec:Kerr-Hall}, the
experimental frequency $\omega$ is in between the upper and lower
plasma frequencies $\omega_{ab}>\omega>\omega_c$ and, thus, $\omega$
can be in resonance with one of the plasma modes (\ref{coll3D2}) for
some vector $\bm q$. Besides, because of the boundary at $z=0$
between the sample and vacuum, it may be necessary to solve for the
plasma modes more accurately by taking into account the surface
plasmons as well.

Another reason for the discrepancy in the magnitude of the Kerr
effect may be the idealization of our theoretical model.  Indeed, we
considered the electromagnetic response of a clean superconductor in
the presence of particle-hole symmetry.  Within this model,
Eqs.~(\ref{Q00_1})--(\ref{Qchiral}) for the response kernels are
written assuming momentum conservation for the electrons. A more
realistic model has to take into account the effect of impurities,
in which case the momentum conservation does not hold.  Indeed, one
of the reasons for smallness of $\theta_K$ within our model can be
traced back to the cancellation of the two terms in
Eq.~(\ref{combination}) at high frequencies.  As a result, the
right-hand side of Eq.~(\ref{combination}) is small in the parameter
$(\Delta_0/\hbar\omega)^2\sim 10^{-6}$.  If we take into account
impurities, the complete cancellation in Eq.~(\ref{combination})
might not hold, and the estimate for $\theta_K$ would be
significantly larger.  Indeed, in the derivation of this equation,
we assumed that the paramagnetic kernel $Q^{(2)}_{kl}$ given in
Eq.~(\ref{Qij_2}) vanishes in the dynamic limit ($\omega\neq0$ and
$\bm q\to 0$). However, it is well known from the classic paper by
Mattis and Bardeen \cite{Mattis-Bardeen} that $Q^{(2)}_{kl}$ does
not vanish when impurity scattering is taken into account.  This
term plays a significant role by taking away some spectral weight
from the London-Meissner supercurrent, which makes the
superconducting gap visible in optical measurements.  For $\rm
Sr_2RuO_4$, this fact was experimentally confirmed in
Ref.~\cite{Ormeno}, which found that the superfluid density at $T=0$
is reduced by $22 \%$.  Therefore, the cancellation in
Eq.~(\ref{combination}) in the presence of disorder should be
investigated in a future theoretical work.

\subsection{Proposed experiments}
\label{Sec:AFM}

\begin{figure} \centering
\includegraphics[height=2.5in]{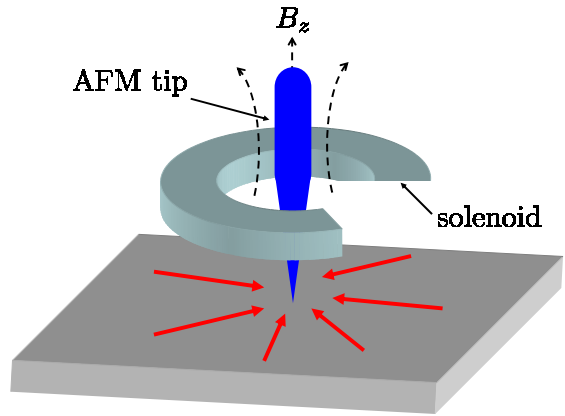}
  \caption{(color online). Schematic setup of a proposed experiment to
  measure the electric charge induced by a magnetic field
  perpendicular to the layers in $\rm Sr_2RuO_4$, as given by Eq.\
  (\ref{charge-response-static}).  A current in the solenoid or the
  coil produces a magnetic field $B_z$ inducing an electric charge at
  the center, which is carried by the radial currents shown by the
  arrows.  The induced charge is detected by the AFM tip.}
\label{newexp}
\end{figure}

As discussed in Sec.~\ref{Sec:charge}, the magnetic field component
$B_z$ applied perpendicular to the conducting planes of $\rm
Sr_2RuO_4$ is expected to induce an electric charge.  The effect is
stronger in the static limit than in the dynamic limit.  In this
section, we propose a conceptual experiment to verify this effect.
The schematic experimental setup is shown in Fig.~\ref{newexp}.  A
miniature solenoid or a coil carrying electric current creates a
magnetic field with the $B_z$ component.  The current and the
magnetic field are modulated at a low frequency (a few kilohertz),
so that the quasistatic formula (\ref{charge-response-static}) is
applicable~\cite{newexp}, but the lock-in measurements are enabled.
The slowly alternating magnetic field $B_z$ induces an alternating
electric charge near the center of the coil, which is produced by
the radial alternating currents, shown by the arrows in
Fig.~\ref{newexp}.  The induced electric charge is detected by a
sensitive atomic force microscope (AFM) tip, shown as the pointed
vertical object in Fig.~\ref{newexp}.  The AFM tip should be made
from a nonmagnetic but highly electrical polarizable material to
reduce the response to magnetic fields and enhance the response to
the electric charge.  To increase sensitivity, measurements should
be performed at the lock-in frequency of the current in the coil.
The difference of the signals above and below the superconducting
transition temperature should be taken in order to subtract the
background effects.  A noticeable increase of the signal below $T_c$
would signify the observation of the effect.  Then, the measurements
can be performed at different locations on the sample.  If a
superconducting domain of the opposite chirality is found, the
effect would change sign, i.e.,\ the sign of the induced electric
charge would change vs the sign of the applied magnetic field.  It
may be possible to map out the chiral domain boundaries in this way,
assuming that the domains are sufficiently static and do not
fluctuate in time too fast.

\section{Conclusions}
\label{Sec:conclusions}

In this paper, we studied the electromagnetic response of a Q2D
chiral $p_x+ip_y$ superconductor, such as $\rm Sr_2RuO_4$.  By
integrating out the superconducting phase $\Phi$, we obtained the
gauge-invariant effective action of the system in an electromagnetic
field.  Besides the well-known conventional terms, this action also
contains an anomalous chiral term, which breaks the time-reversal
symmetry. Instead of the Chern-Simons-type term (\ref{CS}),
discussed in the earlier literature
\cite{Volovik88,Furusaki01,Yakovenko07}, the gauge-invariant
anomalous effective action (\ref{S_a}) contains a product of the
electric and magnetic fields. This is rather unusual situation, and
the modified Chern-Simons-type term (\ref{S_a}) leads to a number of
interesting effects.

By taking variations of the anomalous effective action with respect
to the electromagnetic potentials $A^{\mu}$, we obtained the
anomalous charge and current responses of the system.  We found that
the transverse chiral response couples to the collective plasma
modes, while the anomalous charge response manifests itself as the
electric charge induced by the magnetic field $B_z$ perpendicular to
the conducting layers of $\rm Sr_2RuO_4$.  In Sec.\ \ref{Sec:AFM},
we proposed an experiment for the detection of this charge.  The
effect is stronger in the static case than at high frequencies.

The anomalous current response manifests itself as the intrinsic Hall
effect, i.e.,\ the existence of the antisymmetric Hall conductivity
$\sigma_{xy}=-\sigma_{yx}$ in the absence of an external magnetic
field.  In the static case, it is a formal consequence of the
magnetization current $\bm j=\bm\nabla\times\bm M$ (the Mermin-Muzikar
current \cite{Mermin80}) in the presence of an inhomogeneous electron
density induced by an electric field.  Because the Cooper pairing
takes place with the angular momentum $L_z=\pm\hbar$, the variation of
electron density causes a variation of magnetization. However, the
equilibrium magnetization currents, obtained in the static limit, are
solenoidal and cannot be directly measured in the transport
experiments. In the dynamic limit, our calculations show that the
magnetization current is greatly reduced by the parameter
$(\Delta_0/\omega)^2$.  In addition to the magnetization current,
there is also an anomalous polarization current, which originates from
the motion of the electric charges induced by the magnetic field
$B_z$. Both the magnetization and polarization currents are of the
same order and are suppressed at high frequencies.

We found that the calculated Hall conductivity is proportional to
$q_\|^2$, where $q_\|$ is the in-plane wave vector of the
electromagnetic wave.  This is an interesting result, which follows
from the cancelation of the electric field by the supercurrent. The
latter was not taken into account in the previous papers
\cite{Yakovenko07,Mineev07}.  As a result, we found that the Kerr
angle $\theta_K$ should depend on the transverse size $l$ of the
laser beam (see Fig.\ \ref{fig:setup}) according to Eq.\
(\ref{size}). Within the considered model here (a clean
superconductor in the presence of particle-hole symmetry), we found
that the overall magnitude of the Kerr angle is much smaller than
experimentally observed in Ref.\ \cite{Xia06}.  We pointed out two
possible reasons for this discrepancy: coupling to the plasma
resonances and the effect of impurity scattering. In the presence of
impurities, the quasiparticle momentum is not conserved, and the
results of our calculations would change significantly. We conclude
that, in order to understand the experiment~\cite{Xia06}, it is
necessary to take into account the effect of disorder.

\textit{Note added.} Recently, a preprint by Roy and Kallin
~(Ref.~\cite{RoyKallin}) appeared, in which similar results were
found using a somewhat different approach.

\begin{acknowledgements}
This work was supported by the Joint Quantum Institute Postdoctoral
Fellowship (RML) and Graduate Assistantship (PN).  The authors (RML
and VMY) acknowledge the hospitality of the Kavli Institute for
Theoretical Physics, Santa Barbara (supported by the National
Science Foundation under Grant No.~PHY05-51164), where a part of
this work was done during the Miniprogram ``$\rm Sr_2RuO_4$ and
Chiral $p$-wave Superconductivity.''  We would like to thank
A.~Kapitulnik, C.~Kallin, S.~Das~Sarma, H.~D.~Drew, V.~P.~Mineev,
A.~Goldman C.~Nayak, M.~Stone, R.~Roy, L.~Cywinski, E.~Hwang,
E.~Rossi, S.~Tewari, and J.~Xia for stimulating discussions.
\end{acknowledgements}

\appendix

\section{Polarization functions}
\label{App:Polarization}

In this appendix, we present analytical expressions for the
correlation functions given by Eqs.\ (\ref{Q00})--(\ref{Qj0a}) and
generalized to the 3D case as explained in Sec.\ \ref{Sec:Q2D}.
After calculating the Matsubara sums and performing the analytical
continuation to the real frequency, we find the following
expressions for the finite-temperature causal polarization functions

\begin{widetext}
\begin{eqnarray}
  Q_{00}&=&\!-\frac{e^2}{2}\!\int{\frac{d^3 p}{(2\pi)^3}}\!\left[\left(
  1\!+\!\frac{\xi_{+}\xi_{-} \!-\! \Delta_+
  \Delta_-}{E_{+}E_{-}}\right)\left(
  \frac{1}{E_+\!-\!E_-\!+\!\omega\!+\!i\delta}\!+\!
  \frac{1}{E_{+}\!-\!E_-\!-\!\omega-i\delta}\right)
  [f(E_-)\!-\!f(E_+)]\right.
\nonumber\\
  &+&\left.\left(1\! - \!\frac{\xi_{+}\xi_{-} \!-\!
  \Delta_+ \Delta_-}{E_{+} E_-}\right)
  \left(\frac{1}{E_+\!+\!E_-\!+\!\omega\!+\!i\delta}\!+\!
  \frac{1}{E_{+}\!+\!E_-\!-\!\omega\!-\!i\delta}\right)
  [1\!-\!f(E_-)\!-\!f(E_+)]\right],
\label{Q00_1}
\end{eqnarray}
\begin{eqnarray}
  Q^{(2)}_{kl}\!&\!=\!&\!-\frac{e^2}{2}\!\!\int{\frac{d^3 p}{(2\pi)^3}}
  v_k v_l
  \left[\left(1\!+\!\frac{\xi_{+}\xi_{-} \!+\! \Delta_+
  \Delta_-}{E_{+}E_{-}}\!\right)\!\!\left(\!
  \frac{1}{E_+\!-\!E_-\!+\!\omega\!+\!i\delta}\!+\!
  \frac{1}{E_{+}\!-\!E_-\!-\!\omega\!-\!i\delta}\right)\!
  [f(E_-)\!-\!f(E_+)]\right.
\nonumber\\
  \!&\!+\!&\!\left.\left(1\!-  \!
  \frac{\xi_{+}\xi_{-} \!+\! \Delta_+ \Delta_-}{E_{+} E_-}\right)
  \left(\frac{1}{E_+\!+\!E_-\!+\!\omega\!+\!i\delta}\!+\!
  \frac{1}{E_{+}\!+\!E_-\!-\!\omega\!-\!i\delta}\right)
  [1\!-\!f(E_-)\!-\!f(E_+)]\right],
\label{Qij_2}
\end{eqnarray}
\begin{eqnarray}
  Q^{(s)}_{k0}\!&\!=\!&\!-\frac{e^2}{2}\!\int{\frac{d^3 p}{(2\pi)^3}}
  v_k \left[\left(
  \frac{\xi_{+}}{E_{+}}-\frac{\xi_{-}}{E_{-}}\right)
  \left[\frac{1}{E_+\!+\!E_-\!+\!\omega\!+\!i\delta}\!-\!
  \frac{1}{E_{+}\!+\!E_-\!-\!\omega\!-\!i\delta}\right]
  [1-f(E_-)\!-\!f(E_+)]\right.
\nonumber\\
  &+&\left.\left(\frac{\xi_{+}}{E_{+}}+\frac{\xi_{-}}{E_{-}}\right)
  \left(\frac{1}{E_+\!-\!E_-\!+\!\omega\!+\!i\delta}\!-\!
  \frac{1}{E_{+}\!-\!E_-\!-\!\omega\!-\!i\delta}\right)
  [f(E_-)\!-\!f(E_+)]\right],
\label{Qj0_1}
\end{eqnarray}
\begin{eqnarray}
  Q^{(a)}_{k0}\!&\!=\!&\!\frac{i e^2}{2}\!\int{\frac{d^3 p}{(2\pi)^3}}
  v_k
  \frac{(p_xq_y\!-\!p_yq_x)\Delta_x\Delta_y}{E_+E_-}\!
  \left[\!\left(\frac{1}{E_{+}\!+\!E_{-}\!+\!\omega\!+\!i\delta}\!+\!
  \frac{1}{E_{+}\!+\!E_{-}\!-\!\omega\!-\!i\delta}\right)
  [1\!-\!f(E_-)\!-\!f(E_+)]\right.
\nonumber\\
  &\!-\!&\left.\left(\frac{1}{E_{+}\!-\!E_{-}\!+\!\omega\!+\!i\delta}
  \!+\!\frac{1}{E_{+}\!-\!E_{-}\!-\!\omega\!-\!i\delta}\right)
  [f(E_-)\!-\!f(E_+)]\right].
\label{Qchiral}
\end{eqnarray}
\end{widetext}
The notation is explained after Eq.\ (\ref{Q00}), and the
integration over $p_z$ runs from $-\pi/d$ to $\pi/d$. Here, $f(E)$
is the Fermi distribution function.  The diamagnetic term
$Q^{(1)}_{kj}$ is given by Eq.\ (\ref{Qij13D}).

\section{Collective modes}
\label{App:Collective}

An alternative way of deriving the collective modes~\cite{Sharapov}
is to set $A^{\mu}=0$ in Eq.~(\ref{action1}) and study the dynamics
of the internal scalar potential $\varphi(q)$ coupled with the
superconducting phase $\Phi(q)$.  After an analytical continuation
to the real frequencies, the effective action becomes
\begin{eqnarray}
  && \!\!\!\!\!\!\!\! S_{\rm eff}(\Phi, \varphi)=
\\
  && =\frac{1}{2}\sum_q
  \!\left(\frac{1}{2e}\Phi(q),\varphi(q)\right)\!\left(
  \begin{array}{cc}
    M_{11} & M_{12} \\
    M_{21} & M_{22} \\
  \end{array}
  \right) \!\left(\begin{array}{c}
    \frac{1}{2e}\Phi(-q) \\
    \varphi(-q)\!
  \end{array}\right).
\nonumber
\end{eqnarray}
Here, the matrix elements are
\begin{eqnarray}
  &&\!\!\!\!\!\!\!\! M_{11}=Q_{kl}q_kq_l\!+\!Q_{00}\omega^2
  \!+\!(Q_{0k}+Q_{k0})\omega q_k,
\\
  && \!\!\!\!\!\!\!\! M_{22}\!=\!V(\bm q)^{-1}\!-\!Q_{00}, \quad
  M_{12}\!=\!-\!M_{21}=-\left[Q_{00}\omega\!+\!Q_{k0}q_k\right].
\nonumber
\end{eqnarray}
Dispersion of this mode is determined by setting the determinant of
the matrix $M$ to zero,
\begin{equation}
  \mathrm{Det}(M)=M_{11}M_{22}-M_{21}M_{12}=0.
\end{equation}
This equation is the same as the equation $\tilde R=0$ with $\tilde R$
given by Eq.~(\ref{R3D}).

\section{Alternative derivation of the effective action}
\label{Sec:alt-action}

In this appendix, we give a simplified alternative derivation of the
effective action for a chiral superconductor after elimination of
the superconducting phase $\Phi$.  This derivation is more
transparent and easier to compare with the effective actions
discussed in Refs.\
\cite{Volovik88,Goryo98,Goryo99,Golub03,Stone04}.

The starting point is the effective action (\ref{action1}), which is
explicitly written below in the tensor components
\begin{widetext}
\begin{equation}
  S = \frac12 \sum_q
  Q_{00} \left(A_0-\frac{i\omega}{2e}\Phi\right)
  \left(A_0+\frac{i\omega}{2e}\Phi\right)
  +Q_{kl} \left(A_k-\frac{iq_k}{2e}\Phi\right)
  \left(A_l+\frac{iq_l}{2e}\Phi\right)
  +2Q_{k0} \left(A_0-\frac{i\omega}{2e}\Phi\right)
  \left(A_k+\frac{i
  q_k}{2e}\Phi\right).
\label{action-start}
\end{equation}
\end{widetext}
Here, in the spirit of Sec.\ \ref{Sec:Total-external}, we take the
electromagnetic potentials $A^\mu$ to represent the total
electromagnetic field, without separating into the external and
internal parts, so we drop the terms with the internal scalar
potential $\varphi$ from Eq.\ (\ref{action1}).  To shorten notation,
it is implied in each term of Eq.\ (\ref{action-start}) that the
first dynamical variable has the argument $q$, e.g.,\ $A_0(q)$, and
the second variable has the argument $-q$, e.g.,\ $A_0(-q)$, as in
Eq.\ (\ref{action1}).  The effective action (\ref{action-start}) is
written using the real frequency $\omega$, as discussed in
Sec.~\ref{Sec:anomalous-action}.

The superconducting phase $\Phi$ in Eq.\ (\ref{action-start}) is the
dynamical variable, which should be eliminated by minimizing $S$
with respect to $\Phi$ and integrating it out.  Before doing so, we
shift the variable $\Phi$ by introducing a different variable
$\phi$,
\begin{equation}
  \Phi(q) = \phi(q) + \frac{2e}{i\omega}\,A_0(q).
\label{Phi-phi}
\end{equation}
Substituting Eq.\ (\ref{Phi-phi}) into Eq.\ (\ref{action-start}) and
sorting out the obtained terms, we find
\begin{equation}
  S = \frac12 \sum_q Q_{kl}\frac{E_kE_l}{\omega^2}
  +R\frac{\phi\phi}{(2e)^2}
  +2W_k\frac{\phi}{2e}\frac{E_k}{\omega}
  +2\Theta\frac{\phi}{2e}i\omega B_z,
\label{action-shifted}
\end{equation}
where $E_k$ and $B_z$ are the electric and magnetic fields.  The
first term in Eq.\ (\ref{action-shifted}) corresponds to the Drude
response of a metal at high frequencies, as in Eq.\
(\ref{tensor-sigma}).  The second term in Eq.\
(\ref{action-shifted}) represents the effective action for the
collective variable $\phi$, where the function $R$ is given by Eq.\
(\ref{R}).  The last two terms in Eq.\ (\ref{action-shifted})
represent the interaction between the collective variable $\phi$ and
the electromagnetic field.  The third term is the conventional one,
where the function $W_k$ is given by Eq.\ (\ref{W}). The last term
corresponds to the anomalous Chern-Simons-type term (\ref{CS}) with
the function $\Theta$ being given by Eq.~(\ref{Theta'}). Equation
(\ref{action-shifted}) is manifestly gauge-invariant, because it
contains only the electric and magnetic fields, rather than the
scalar and vector potentials.

The effective action (\ref{action-shifted}) is quadratic with
respect to $\phi$.  After the elimination of $\phi$, we obtain the
final result
\begin{eqnarray}
  S &=& \frac12 \sum\limits_q \left(Q_{kl}-\frac{W_kW_l}{R}\right)
  \frac{E_k(q)E_l(-q)}{\omega^2}
\label{action-final} \\
  && {}- 2i\frac{\Theta W_k}{R}E_k(q)B_z(-q)
  - \frac{\omega^2\Theta^2}{R}B_z(q)B_z(-q).
\nonumber
\end{eqnarray}
The first term in Eq.\ (\ref{action-final}) represents the
conventional contribution to the effective action discussed in
Sec.~\ref{Sec:Total-conventional}.  The second term is the chiral
anomalous term (\ref{S_a}) representing the modification of the
Chern-Simons-type term due to the dynamics of the collective
variable $\phi$.  The third term in Eq.\ (\ref{action-final}) is the
double-anomalous nonchiral term, which was briefly mentioned at the
beginning of Sec.~\ref{Sec:Anomalous-response}.  These terms are
discussed in more detail below.

The effective actions similar to Eq.\ (\ref{action-final}) were
obtained for chiral superconductors after integration out of the
superconducting phase in Refs.\ \cite{Goryo98,Goryo99,Golub03}.
However, the focus in these papers was on the low-frequency limit,
and some terms in the effective action were omitted in this limit.
The high-frequency limit of the effective action, which is relevant
for the experiment \cite{Xia06}, was not discussed in literature
before, except for Ref.\ \cite{Yakovenko07}.  It is remarkable that,
in the high-frequency limit, all three different functions $\Theta$,
$\bm W$, and $R\approx Q_{00}\omega^2$ are proportional to the same
function $I(\omega)$,
\begin{equation}
  \Theta(\omega), \: \bm W(\omega), \: Q_{00}(\omega) \:
  \propto \: I(\omega) \: \propto \: \frac{\Delta_0^2}{\omega^2},
\label{high-w}
\end{equation}
as follows from Eqs.\ (\ref{Theta'}), (\ref{combination}),
(\ref{Q_00-w}), and (\ref{asymptotes}).  As a consequence of Eq.\
(\ref{high-w}), the last terms in Eqs.\ (\ref{action-final}) are
proportional to $I(\omega)$ and tend to zero as
$(\Delta_0/\omega)^2$ at high frequencies.  Only the very first term
in Eq.\ (\ref{action-final}) survives, which represents the Drude
response of free electrons at high frequencies.  The suppression of
the last terms in Eq.~(\ref{action-final}) at high frequencies
should be expected, because they represent the effect of the chiral
superconducting state. In the limit $\Delta_0/\omega\to0$, the
superconducting effects should vanish, because the low-energy gap
$\Delta_0$ cannot affect the high-energy behavior, as discussed in
Sec.~\ref{Sec:magnetization}.

By using Eq.\ (\ref{high-w}), the last double-anomalous term in
Eq.~(\ref{action-final}) can be written as
\begin{equation}
  S^{(aa)} = \frac{e^2}{2\pi dm_ec^2} \sum_q I \,
  \frac{B_z(q)\,B_z(-q)}{8\pi},
\label{S_aa'}
\end{equation}
where we restored the dimensional units.  Comparing with the action
of a free magnetic field in vacuum $\sum_q \bm B(q)\cdot\bm
B(-q)/8\pi$, we see that Eq.~(\ref{S_aa'}) gives an orbital
paramagnetic contribution to the magnetic susceptibility of the
system.  This contribution is small because of the relativistic
factor $e^2/dm_ec^2\ll1$ in Eq.\ (\ref{S_aa'}).  Given that this
term is nonchiral and small, we do not pay much attention to the
double-anomalous term in the paper.

The first term in Eq.~(\ref{action-final}), which involves the product
$E_kE_l$, represents the dielectric susceptibility of the system.  For
small $\bm q$, the term proportional to $W_kW_l$ can be omitted.  The
remaining term proportional to $Q_{kl}$, when combined with the action
of a free electric field in vacuum $-\sum_q \bm E(q)\cdot\bm
E(-q)/8\pi$, gives the the conventional effective action for the
electric field in the medium
\begin{equation}
  S^{(c)}_E = -\frac{1}{8\pi} \sum_q
  \bm E(q) \cdot\tensor{\epsilon} \cdot\bm E(-q),
\label{S_E}
\end{equation}
where the dielectric permeability tensor $\tensor{\epsilon}$ is
given by Eq.~(\ref{tensor-epsilon}).  If we substitute the internal
electric field in the form $\bm E^{\rm int}=i\bm q\varphi$ into
Eq.~(\ref{S_E}), then we can obtain the equation $\bm
q\cdot\tensor{\epsilon}\cdot\bm q=0$ for the spectrum of the
collective plasma modes, which is the same as Eq.~(\ref{coll3D2}).

The effective action (\ref{action-final}) is manifestly gauge
invariant, and the charge and current response functions, discussed
in the rest of the paper, can be obtained by taking a variation of
this action.  The function $R$ (\ref{R}) in the denominators of Eq.\
(\ref{action-final}) has zeros at the frequencies of the collective
modes of $\Phi$ \cite{acoustic}.  If it is desirable to directly
include the internal Coulomb interaction in the effective action
(rather than by using Maxwell's equations), the kernels $Q$ should
be replaced by the kernels $\tilde Q$ (\ref{tilde_Q}) in Eq.\
(\ref{action-start}) and $R$ should be replaced by $\tilde R$ in
Eq.\ (\ref{action-final}).

\section{Relationship between the polar Kerr angle and ac Hall
conductivity}
\label{Sec:Kerr-Hall}

In this appendix, we obtain the equations expressing the polar Kerr
angle in terms of the ac Hall conductivity~(see also
Ref.~\cite{Mineev07}). To simplify the notation, we drop the index
$(a)$ from $\sigma_{xy}^{(a)}$.

Let us consider the normal reflection of a linearly polarized
electromagnetic plane wave, propagating in the $z$ direction, from
the $(x,y)$ surface of a Q2D chiral superconductor.  The reflection
coefficient $|r|$ and the polar Kerr angle $\theta_K$ are given by
the following equations \cite{White-Geballe}
\begin{eqnarray}
  && |r|=\frac{|n-1|}{|n+1|},
\label{r} \\
  && \theta_K=\frac{4\pi}{\omega}\,{\rm Im}\frac{\sigma_{xy}}{n\,(n^2-1)},
\label{theta_K}
\end{eqnarray}
where $n$ is the complex refraction coefficient.  The refraction
coefficient is obtained from the dielectric susceptibility tensor,
which has the general form
\begin{equation}
  \tensor{\epsilon} = \tensor{\epsilon}^{(\infty)}
  +\frac{4\pi i}{\omega}\tensor{\sigma}.
\label{epsilon-infty}
\end{equation}
Here, the conductivity tensor $\tensor{\sigma}$ is given by
Eq.~(\ref{tensor-sigma}), and $\tensor{\epsilon}^{(\infty)}$ is the
background dielectric tensor, which originates from the
polarizability of the other, nonconducting bands in the material.
For a plane wave with the electric field polarized in the $x$
direction, the appropriate component of the conductivity tensor is
$\sigma_{xx}$. Then, the corresponding refraction coefficient in
Eqs.\ (\ref{r}) and (\ref{theta_K}) is
\begin{equation}
  n^2=\epsilon_{xx}=\epsilon_\infty-\frac{\omega_{ab}^2}{\omega^2},
  \qquad  \epsilon_\infty\equiv\epsilon^{(\infty)}_{xx}\geq1,
\label{n}
\end{equation}
where we introduced a shorthand notation $\epsilon_\infty$. Equation
(\ref{tensor-sigma}), derived for an ideal clean superconductor,
gives only the reactive imaginary part $\sigma_{xx}''$ of
conductivity and does not contain the dissipative, real part
$\sigma_{xx}'$.  To keep the presentation simple, here we discuss
only the ideal case with $\sigma_{xx}'=0$, but the consideration can
be generalized to include $\sigma_{xx}'\neq0$. Notice also that
Eq.~(\ref{theta_K}) was derived assuming that the off-diagonal
component $\sigma_{xy}$ of the conductivity tensor is much smaller
than the diagonal component $\sigma_{xx}$.

The refraction coefficient $n$ in Eq.~(\ref{n}) vanishes at
$\omega=\omega_p$, where $\omega_p$ is the so-called plasma edge
frequency
\begin{equation}
  \omega_p=\omega_{ab}/\sqrt{\epsilon_\infty}.
\label{w_p}
\end{equation}
Equations (\ref{r}), (\ref{theta_K}), and (\ref{n}) have different
forms for the frequencies above and below the plasma edge.  For
$\omega>\omega_p$, the refraction coefficient $n$ in Eq.~(\ref{n}) is
real, so the reflection coefficient (\ref{r}) is $0<|r|<1$, i.e.,\ the
electromagnetic wave is partially reflected and partially transmitted
through the crystal.  The polar Kerr angle (\ref{theta_K}) is given by
the following expression in this case
\begin{equation}
  \theta_K=\frac{4\pi\omega^2\,\sigma_{xy}''}
  {\sqrt{\epsilon_\infty\omega^2-\omega_{ab}^2} \,
  [(\epsilon_\infty-1)\omega^2-\omega_{ab}^2]},
  \quad \omega>\omega_p.
\label{theta_K-high}
\end{equation}
For frequencies $\omega<\omega_p$ below the plasma edge, the
refraction coefficient $n$ (\ref{n}) is imaginary.  In this case, the
electromagnetic wave is completely reflected ($|r|=1$), and the polar
Kerr angle (\ref{theta_K}) is
\begin{equation}
  \theta_K=-\frac{4\pi\omega^2\,\sigma_{xy}'}
  {\sqrt{\omega_{ab}^2-\epsilon_\infty\omega^2} \,
  [(\epsilon_\infty-1)\omega^2-\omega_{ab}^2]}, \quad \omega<\omega_p.
\label{theta_K-low}
\end{equation}

Equations (\ref{theta_K-high}) and (\ref{theta_K-low}) show that the
Kerr angle is determined by the imaginary part $\sigma_{xy}''$ for
$\omega>\omega_p$ and by the real part $\sigma_{xy}'$ for
$\omega<\omega_p$.  This statement becomes approximate when the
dissipative component $\sigma_{xx}'$ is taken into account, and both
the real and imaginary parts of $\sigma_{xy}$ start to contribute
simultaneously to $\theta_K$.  In our theory, $\sigma_{xy}'$ and
$\sigma_{xy}''$ are proportional to the real and imaginary parts of
the function $I(\omega)$ defined in Eqs.\ (\ref{Theta'}) and
(\ref{I}).  The plots of the real and imaginary parts of $I(\omega)$
are shown in Fig.\ \ref{figI1}, and their asymptotic expressions are
given by Eq.\ (\ref{asymptotes}) \cite{I(omega)}.

One can notice that the denominators of Eqs.~(\ref{theta_K-high})
and (\ref{theta_K-low}) vanish at certain frequencies, providing
resonance enhancement of the Kerr angle $\theta_K$.  At the plasma
edge frequency $\omega=\omega_p$, the square roots in the
denominators of Eqs. (\ref{theta_K-high}) and (\ref{theta_K-low})
vanish.  At the frequency
$\omega=\omega_{ab}/\sqrt{\epsilon_\infty-1}>\omega_p$, the
denominator of (\ref{theta_K-high}) vanishes, and $\theta_K$ changes
sign.  At this frequency, the reflection coefficient (\ref{r})
vanishes because $n=1$ in Eq.\ (\ref{n}).  Of course, in the
presence of $\sigma_{xx}'$, these singularities will be smeared out.

The optical properties of $\rm Sr_2RuO_4$ in the normal state were
measured in Refs.\ \cite{Tokura96,Tokura01}; however, the main
interest was in the electric field polarization along the $\bm c$
axis.  The plasma frequencies $\omega_{ab}=4.5$ eV and
$\omega_c=0.32$ eV were obtained from the fits of the data, as well
as the relaxation rate $\gamma\sim0.5$ eV.  The value
$\epsilon_\infty=10$ was quoted, but for $\bm E\|\bm c$.  In the
Kerr effect measurements \cite{Xia06}, the frequency of the incoming
light was $\omega=0.8$ eV.  This frequency is clearly below the
in-plane plasma frequency $\omega_{ab}$ and probably lower than
$\omega_p$, even for $\epsilon_\infty=10$.  In this limit, the Kerr
angle is given by Eq.~(\ref{theta_K-low}). However, the effect of
the relaxation rate $\gamma$, which is comparable with $\omega$,
should also be taken into account.



\begin{thebibliography}{99}

\bibitem{Rice} M. Rice, Science {\bf 314}, 1248 (2006).

\bibitem{Day} C. Day, Phys. Today {\bf 59(12)}, 23 (2006).

\bibitem{Xia06} J.~Xia, Y.~Maeno, P.T.~Beyersdorf, M.M.~Fejer, and
  A.~Kapitulnik, Phys. Rev. Lett. {\bf 97}, 167002 (2006).

\bibitem{Kidwingira06} F.~Kidwingira, J.D.~Strand, D.J.~Van Harlingen,
  and Y.~Maeno, Science {\bf 314}, 1267 (2006).

\bibitem{Luke98} G.M.~Luke, Y.~Fudamoto, K.M.~Kojima, M.I.~Larkin,
  J.~Merrin, B.~Nachumi, Y.J.~Uemura, Y.~Maeno, Z.Q.~Mao, Y.~Mori,
  H.~Nakamura, and M.~Sigrist, Nature~(London) {\bf 394}, 558 (1998).

\bibitem{Luke00} G.M.~Luke, Y.~Fudamoto, K.M.~Kojima, M.I.~Larkin,
  B.~Nachumi, Y.J.~Uemura, J.E.~Sonier, Y.~Maeno, Z.Q.~Mao, Y.~Mori,
  and D.F.~Agterberg, Physica B {\bf 289--290}, 373 (2000).

\bibitem{quarter-wave} Technically, the experiment \cite{Xia06} also
  included a quarter-wave plate, which converted the linear
  polarization into the circular one and back to linear after
  reflection.  This is not essential for our theoretical
  consideration, because the electromagnetic response can be studied
  in any polarization basis (linear or circular) and then an
  appropriate superposition can be taken.

\bibitem{Moller05} P.G.~Bjornsson, Y.~Maeno, M.E.~Huber, and
  K.A.~Moler, Phys. Rev. B {\bf 72}, 012504 (2005).

\bibitem{Moller07} J.R.~Kirtley, C.~Kallin, C.W.~Hicks, E.-A.~Kim,
  Y.~Liu, K.A.~Moler, Y.~Maeno, and K.D.~Nelson, Phys. Rev. B {\bf
  76}, 014526 (2007).

\bibitem{Mackenzie03} A.~P.~Mackenzie and Y.~Maeno,
  Rev. Mod. Phys. {\bf 75}, 657 (2003).

\bibitem{Bergemann03} C.~Bergemann, A.P.~Mackenzie, S.R.~Julian,
  D.~Forsythe, and E.~Ohmichi, Adv. Phys. {\bf 52}, 639 (2003).

\bibitem{Baskaran96} G.~Baskaran, Physica B {\bf 223--224}, 490
  (1996).

\bibitem{Rice95} T.M.~Rice and M.~Sigrist, J. Phys. Condens. Matter {\bf
  7}, L643 (1995).

\bibitem{Volovik88} G.E.~Volovik, Sov. Phys. JETP {\bf 67}, 1804
  (1988).

\bibitem{Mazin05} I. \u{Z}uti\'c and I. Mazin, Phys. Rev. Lett. {\bf
  95}, 217004 (2005).

\bibitem{Ishida98} K.~Ishida, H.~Mukuda, Y.~Kitaoka, K.~Asayama,
  Z.Q.~Mao, Y.~Mori, and Y.~Maeno, Nature~(London) {\bf 396}, 658 (1998).

\bibitem{Ishida01} K.~Ishida, H.~Mukuda, Y.~Kitaoka, Z.Q.~Mao,
  H.~Fukazawa, and Y.~Maeno, Phys. Rev. B {\bf 63}, 060507(R) (2001).

\bibitem{Murakawa04} H.~Murakawa, K.~Ishida, K.~Kitagawa, Z.Q.~Mao,
  and Y.~Maeno, Phys. Rev. Lett. {\bf 93}, 167004 (2004).

\bibitem{Nelson04} K.~D.~Nelson, Z.~Q.~Mao, Y.~Maeno, and Y.~Liu,
  Science {\bf 306}, 1151 (2004).

\bibitem{Sengupta02} K.~Sengupta, H.-J.~Kwon, and V.M.~Yakovenko,
  Phys. Rev. B {\bf 65}, 104504 (2002).

\bibitem{White-Geballe} R.M.~White and T.H.~Geballe, \emph{Long Range
Order in Solids} (Academic, New York, 1979), pp. 317, 321.

\bibitem{Joynt91} Q.P.~Li and R.~Joynt, Phys. Rev. B {\bf 44}, 4720
  (1991).

\bibitem{Ting07} W.~Kim et al. [arXiv:0711.0208~(unpublished)]
studied the case where the $p_x$ and $p_y$ components of the
superconducting order parameter are mixed with an arbitrary relative
phase.  They found that $\sigma_{xy}$ is maximal for the real order
parameter $p_x\pm p_y$ and vanishes for the chiral one $p_x\pm
ip_y$. However, their calculated conductivity tensor is symmetric
$\sigma_{xy}=\sigma_{yx}$, so it does not represent the
antisymmetric Hall conductivity ($\sigma_{xy}=-\sigma_{yx}$).  The
real order parameter $p_x+p_y$ does not break the TRS, but breaks
rotational symmetry and creates a preferred direction $\hat {\bm
x}+\hat {\bm y}$.  The conductivity tensor becomes anisotropic and
acquires symmetric off-diagonal components
$\sigma_{xy}=\sigma_{yx}$, which, however, does not contribute to
the PKE in the experiment~(Ref.~\cite{Xia06}).

\bibitem{Yip92} S.K.~Yip and J.A.~Sauls, J. Low Temp. Phys. {\bf 86},
  257 (1992).

\bibitem{Goryo98} J. Goryo and K. Ishikawa, Phys. Lett. A {\bf 246},
  549 (1998).

\bibitem{Goryo99} J. Goryo and K. Ishikawa, Phys. Lett. A {\bf 260},
  294 (1999).

\bibitem{Goryo00} J. Goryo, Phys. Rev. B {\bf 61}, 4222 (2000).

\bibitem{Golub03} B. Horovitz and A. Golub, Phys. Rev. B {\bf 68},
  214503 (2003).

\bibitem{Stone04} M. Stone and R. Roy, Phys. Rev. B {\bf 69}, 184511
  (2004).

\bibitem{Furusaki01} A.~Furusaki, M.~Matsumoto, and M.~Sigrist,
  Phys. Rev. B {\bf 64}, 054514 (2001).

\bibitem{Yakovenko07} V.M.~Yakovenko, Phys. Rev. Lett.  {\bf 98},
  087003 (2007).

\bibitem{Mermin80} N.D.~Mermin and P.~Muzikar, Phys. Rev. B {\bf 21},
  980 (1980).

\bibitem{Mineev07} V.P.~Mineev, Phys. Rev. B {\bf 76}, 212501 (2007).

\bibitem{Kallin} C.~Kallin (private communication).

\bibitem{FISDW} A similar cancellation happens for the quantum Hall
effect (QHE) in the magnetic-field-induced spin-density-wave (FISDW)
state in quasi-one-dimensional conductors~ (Ref.~\cite{Yakovenko96}
and ~\cite{Goan98}). When FISDW is pinned and acts as a static
potential, it produces the QHE. However, if the FISDW is depinned
and is free to slide along the chains, its motion cancels out the
QHE. The difference between the FISDW and superconductivity is that
the FISDW phase (which represents its translational position) is
normally pinned, whereas the superconducting phase $\Phi$ is not
pinned.

\bibitem{Yakovenko96} V.M.~Yakovenko and H.-S.~Goan, J. Phys. I (France)
  {\bf 6}, 1917 (1996); V.M.~Yakovenko, arXiv:cond-mat/0605750~(unpublished).

\bibitem{Goan98} V.M.~Yakovenko and H.-S.~Goan, Phys. Rev. B {\bf
  58}, 10648 (1998).

\bibitem{AES} U.~Eckern, G.~Sch\"{o}n, and V.~Ambegaokar, Phys. Rev. B
  {\bf 30}, 6419 (1984).

\bibitem{Otterlo} A.~van~Otterlo, D.S.~Golubev, A.D.~Zaikin, and
  G.~Blatter, Eur. Phys. J. B {\bf 10}, 131 (1998).

\bibitem{Sharapov} S.G.~Sharapov, H.~Beck, and V.M.~Loktev,
  Phys. Rev. B {\bf 64}, 134519 (2001).

\bibitem{Paramekanti} A.~Paramekanti, M.~Randeria, T.V.~Ramakrishnan,
  and S.S.~Mandal, Phys. Rev. B {\bf 62}, 6786 (2000).

\bibitem{Mahan} G.D.~Mahan, \emph{Many-Particle Physics}, 2nd ed. (Plenum, New York, 1990).

\bibitem{ReadGreen} N.~Read and D.~Green, Phys. Rev. B {\bf 61}, 10267
  (2000).

\bibitem{Leggett75} A.J.~Leggett, Rev. Mod. Phys. {\bf 47}, 331
  (1975); {\bf 48}, 357 (1976).

\bibitem{Schrieffer} J.R.~Schrieffer, \textit{Theory of
Superconductivity} (Perseus, Reading, MA, 1999).

\bibitem{UFN} P.I.~Arseev, S.O.~Loiko, and N.K.~Fedorov,
  Phys. Usp. {\bf 49}, 1 (2006).

\bibitem{Kulik} I.O.~Kulik, O.~Entin-Wohlman, and R.~Orbach,
  J. Low. Temp. Phys. {\bf 43}, 591 (1981).

\bibitem{Levin} Y.~Zha, K.~Levin, and D.Z.~Liu, Phys. Rev. B {\bf 51},
  6602 (1995).

\bibitem{Das Sarma} H.A.~Fertig and S.~Das Sarma,
  Phys. Rev. Lett. {\bf 65}, 1482 (1990); E.H.~Hwang and S.~Das Sarma,
  Phys. Rev. B {\bf 52}, R7010 (1995).

\bibitem{Negele} J.W.~Negele and H.~Orland, \emph{Quantum
  Many-Particle Systems} (Perseus, Reading, MA, 1998).

\bibitem{Svidzinskii} A.V.~Svidzinskii, \emph{Space-Inhomogeneous
  Problems of Superconductivity} (Nauka, Moscow, 1982).

\bibitem{Euclidean} The Euclidean structure of the
  imaginary-time action (\ref{partition_with_A}) suggests the
  substitution $A_0(t,\bm r)\rightarrow iA_0(\tau,\bm r)$ for the
  time-like component of the electromagnetic field in Eqs.\
  (\ref{L_bos}) and (\ref{L_el}).  As a consequence, the electric
  field is transformed as $\bm E(t,\bm r)\rightarrow i\bm E(\tau,\bm
  r)$, whereas the magnetic field $\bm B$ does not change.  After
  the evaluation of the functional integrals and analytical
continuation
  from the Matsubara to real frequencies, we transform $A_0$ and $\bm
  E$ back.  As a result, the factors $i$ in Eqs.~(\ref{action1}) and
  (\ref{action2}) disappear.

\bibitem{Thouless} P.~Ao, D.J.~Thouless, and X.-M.~Zhu,
  Mod. Phys. Lett. B \textbf{9}, 755 (1995).

\bibitem{Gamma_2} Strictly speaking, the last term in
  Eq.~(\ref{lagrangian2}) acquires the shown momentum structure only
  after averaging over $\psi$.  The momentum structure of this term
  should be understood symbolically.

\bibitem{Kosztin} I.~Kosztin, Q.~Chen, Y.-J.~Kao, and K.~Levin,
  Phys. Rev. B {\bf 61}, 11662 (2000).

\bibitem{Hirschfeld1993} P.J.~Hirschfeld and D.~Einzel, Phys. Rev. B
  {\bf 47}, 8837 (1993).

\bibitem{HwangDasSarma98} E.~Hwang and S.~Das Sarma,
  Int. J. Mod. Phys. B {\bf 12}, 2769 (1998).

\bibitem{Tokura96} T.~Katsufuji, M.~Kasai, and Y.~Tokura,
  Phys. Rev. Lett. {\bf 76}, 126 (1996).

\bibitem{Tokura01} M.G.~Hildebrand, M.~Reedyk, T.~Katsufuji, and
  Y.~Tokura, Phys. Rev. Lett. {\bf 87}, 227002 (2001).

\bibitem{Goldman} R.V.~Carlson and A.M.~Goldman, Phys. Rev. Lett. {\bf
  31}, 880 (1973).

\bibitem{Landau} L.D.~Landau and E.M.~Lifshitz, \emph{Electrodynamics
  of Continuous Media}, 2nd ed. (Pergamon, Oxford, 1984).

\bibitem{I(omega)} The asymptotic expressions (\ref{asymptotes}) for
  the real and imaginary parts of $\Theta(\omega)$ are the same as
  obtained in Eqs.\ (21) and (23) of Ref.\ \cite{Yakovenko07}.  The
  first line of the latter equation gives the real part of
  $\Theta(\omega)$ when the relaxation rate is set to zero:
  $\gamma\to0$.

\bibitem{Ishikawa98} J.~Goryo and K.~Ishikawa, J. Phys. Soc. Jpn.
  {\bf 67}, 3006 (1998).

\bibitem{acoustic} The equation $R(\omega,\bm q)=0$ gives the
  acoustic dispersion relation for the collective modes of the
  superconducting phase $\Phi$ without taking into account the
  internal Coulomb interaction.  Indeed, in the limit of small $\bm q$
  and small $\omega$, using Eqs.\ (\ref{N_0}) and (\ref{tensor-n}) and
  neglecting $Q_{kl}^{(2)}$ and $Q_{0k}^{(s)}$, we find that
  $R(\omega,\bm q)\propto[\omega^2-(v_F^2/2)(\bm q\cdot\tensor{n}\cdot\bm q)]$.

\bibitem{dcHall} A careful calculation using the chiral
  Ginzburg-Landau theory~(Ref.~\cite{Furusaki01}) for the bulk and for the
  edges shows that there may be a very small non-zero dc Hall effect
  in a $p_x+ip_y$ superconductor.

\bibitem{charge} The anomalous induced charge $\delta\rho^{(a)}$ in
  Eq.\ (\ref{charge-response}) is different from the anomalous induced
  charge $\delta\tilde\rho^{(a)}$ in Eqs.\ (\ref{charge-a}) and
  (\ref{polarization-a}).  The latter was obtained by taking a
  variation with respect to $A_0^{\rm ext}$, as opposed to $A_0^{\rm
  tot}$.  The charge $\delta\rho^{(a)}$ is the unscreened charge that
  appears as a source in the right-hand side of Maxwell's equations,
  whereas the charge $\delta\tilde\rho^{(a)}$ is the screened
  anomalous charge, which takes into account the conventional
  screening mechanism.

\bibitem{Streda} P.~\u{S}treda, J. Phys. C {\bf 15}, L717 (1982).

\bibitem{Meschede} D.~Meschede, \emph{Optics, Light and Lasers}, 2nd
ed. (Wiley, Weinheim, 1990).

\bibitem{Mattis-Bardeen} D.C.~Mattis and J.~Bardeen, Phys. Rev. {\bf
  111}, 412 (1958).

\bibitem{Ormeno} R.J.~Ormeno, M.A.~Hein, T.L.~Barraclough, A.~Sibley,
  C.E.~Gough, Z.Q.~Mao, S.~Nishizaki, and Y.~Maeno, Phys. Rev. B {\bf
  74}, 092504 (2006).

\bibitem{newexp} Expression~(\ref{charge-response-static}) does
  not include the effects of electrostatic screening. Therefore, it
  should be understood as an upper limit for the charge estimate.
  However, at the surface of a superconductor electrostatic screening
  is less effective than in the bulk, and this estimate should provide
  a reasonable magnitude for the effect.

\bibitem{RoyKallin} R.~Roy and C.~Kallin,
arXiv.org:0802.3693~(unpublished).


\end{thebibliography}
\end{document}